\documentclass[superscriptaddress,preprintnumbers,footinbib,pra,twocolumn]{revtex4-2}
\usepackage{amsmath}
\usepackage{amssymb}
\usepackage{graphicx}
\usepackage[version=3]{mhchem}
\usepackage{graphicx}
\usepackage{dcolumn}
\usepackage{bm}
\usepackage{hyperref}
\hypersetup{
    colorlinks=true,
    linkcolor=blue,
    filecolor=blue,
    urlcolor=blue,
   citecolor=blue,
}
\usepackage{enumitem}
\usepackage{siunitx}
\usepackage{braket}

\newcommand{\Tr}{\mathop{\mathrm{Tr}}}

\usepackage{xcolor}
\DeclareMathAlphabet\mathbfcal{OMS}{cmsy}{b}{n}
 \usepackage[OMLmathbf]{isomath}
 \allowdisplaybreaks

\begin{document}
\title{Cavity-induced charge transfer in periodic systems: length-gauge formalism}

\author{Ekaterina Vlasiuk}
\affiliation{Department of Physics, University of Basel, Klingelbergstrasse 82, CH-4056 Basel, Switzerland}
\author{Valerii K. Kozin}
\email[Corresponding author: ]{kozin.valera@gmail.com}
\affiliation{Department of Physics, University of Basel, Klingelbergstrasse 82, CH-4056 Basel, Switzerland}
\author{Jelena Klinovaja}
\affiliation{Department of Physics, University of Basel, Klingelbergstrasse 82, CH-4056 Basel, Switzerland}
\author{Daniel Loss}
\affiliation{Department of Physics, University of Basel, Klingelbergstrasse 82, CH-4056 Basel, Switzerland}
\author{Ivan V. Iorsh}
\affiliation{Departament of Physics, Bar-Ilan University, Ramat Gan, Israel}
\author{Ilya V. Tokatly}
\affiliation{IKERBASQUE, Basque Foundation for Science, 48009 Bilbao, Basque Country, Spain}
\affiliation{Donostia International Physics Center (DIPC), 20018 Donostia-San Sebastian, Basque Country, Spain}
\affiliation{Nano-Bio Spectroscopy Group and European Theoretical Spectroscopy Facility (ETSF), Departamento de Polímeros
y Materiales Avanzados: Física, Química y Tecnología, Universidad del País Vasco, 20018 Donostia-San Sebastián,
Basque Country, Spain}


\begin{abstract} 
We develop a length-gauge formalism for treating one-dimensional periodic lattice systems in the presence of a photon cavity inducing light-matter interaction. 
The purpose of the formalism is to remove mathematical ambiguities that occur when defining the position operator in the context of the Power-Zienau-Woolley Hamiltonian. We then use a diagrammatic approach to analyze perturbatively
the interaction between an electronic quantum system and a  photonic cavity mode of long wavelength. We illustrate the versatility of the formalism by studying the cavity-induced electric charge imbalance and polarization in the Rice-Mele model with broken inversion symmetry.
\end{abstract}

\maketitle

\section{Introduction}
Controlling quantum materials and engineering new phases of matter through a light-matter interaction is one of the most promising research directions in condensed matter physics~\cite{cavQuantMatReview}. In the past, light has been mainly used as a tool for probing various properties of quantum systems, such as optical conductivity, from which one may extract a huge amount of useful information about the intricate quantum properties of the material. However, recently the research efforts shifted from using light as a probe to using light as a means of control~\cite{FloquetReview, Bloch2022}.

Over recent years, there has been tremendous interest in inducing novel properties in electronic systems by light, among which it is worth noting the Floquet topological insulator~\cite{PhysRevB.81.165433,PhysRevB.79.081406,Lindner2011,Dehghani2015}, Floquet topological superconductors~\cite{PhysRevB.95.155407,PhysRevLett.116.176401,PhysRevB.88.155133,PhysRevLett.111.136402}, Floquet-engineered topological band structures~\cite{doi:10.1126/science.1239834,McIver2020,PhysRevB.97.035416} in solid-state systems and ultra-cold atoms~\cite{RevModPhys.91.015005,Roux2020,PhysRevLett.118.073602}. The most interesting phenomena usually demand strong light-matter coupling~\cite{FriskKockum2019}, which can be achieved either by increasing the intensity of  light or by confining  light within a small volume inside a cavity~\cite{Maissen2014}. With cavities, one may explore the interaction between matter and vacuum fluctuations of the resonator paving the way for engineering~\cite{PhysRevX.5.031001,PhysRevB.91.085406,PhysRevLett.109.257002,PhysRevResearch.2.043264,Hubener2021,PhysRevB.105.205424,PhysRevX.10.041027,Curtis2023,PhysRevB.102.115310,RiccoKozin2022}  systems in thermodynamic equilibrium in the absence of any external driving, such as, for instance, the modification of the quantum Hall~\cite{FaistHallBreakdown} and the anomalous Hall~\cite{PhysRevB.99.235156} responses via resonators.

In the present work, we focus on one-dimensional quantum systems placed in cavities and our aim is to develop a length-gauge formalism for studying the charge transfer induced by the cavity. The main obstacle on this route is the absence~\cite{Aversa1995} of a well-defined position operator via which the matter couples to light in this gauge. The use of the length gauge has a clear advantage in that it  
uses a gauge-invariant observable, the electric field~\cite{Scully1987}. However, in this gauge, the perturbation is no longer diagonal in momentum. Additionally, the position operator is singular in momentum space.

This is a long-standing problem~\cite{Peres2017,Peres2018}, and a number of attempts to resolve it have been made in the past. For instance, as one of the possible ad-hoc solutions, it was suggested to restrict the application of the coordinate operator, as defined in Eq.~\eqref{CO_Bluont} below, exclusively within commutators. Also, the issue with the position operator was addressed in the context of non-linear response theory where the electromagnetic field is treated classically~\cite{PhysRevB.72.045223,PhysRevB.91.235320,PhysRevB.94.045434,PhysRevB.90.245423,PhysRevB.53.10751}. In this work, we present a step towards the complete solution of this problem for a quantized electromagnetic field. The key idea of our approach 
is to introduce a non-uniformity in the cavity-mode profile. We then exploit the  non-commutativity of taking the uniformity limit 
(rendering the mode uniform) 
either at the beginning or at the end of the calculation that is based on diagrammatic perturbation theory.

To illustrate the usefulness of our formalism we apply it to the calculation of the cavity-induced charge transfer leading to charge imbalance and spontaneous polarization in the ground state. The latter is calculated using the approach suggested by Nourafkan and Kotliar in 2013~\cite{Kotliar2013}. The similarities and discrepancies between the two physical quantities are addressed. 

The work is organized as follows. In Sec. II, we start with performing the gauge transformation to switch from the velocity gauge to the length gauge~\cite{Rokaj2018}. In Sec. III, we introduce a spatial dependence in the vector potential to make the cavity mode non-uniform and generalize the gauge transformation from the previous section. In Sec. IV, we develop a diagrammatic technique in the length gauge and apply it to study the charge transfer in the Rice-Mele model, coupled to a cavity. The main results of the paper are highlighted in the conclusion section. 

\section{Electron-photon Hamiltonian with a constant vector potential}

Our aim is to describe a sufficiently large periodic system, such as a crystallite, a flake of 2D material, or a long quasi-1D chain/nanotube, embedded in a microcavity that supports a discrete set of quantum electromagnetic modes. We consider a typical situation when, on the one hand, the system is much larger than the crystal unit cell, and, on the other hand, it is much smaller than the wavelength of relevant cavity modes. The latter condition justifies the use of the dipole approximation we are using in the present work, which assumes that the mode electric field is practically uniform throughout the system. If the former condition is fulfilled the bulk properties of the system should be independent of the boundary conditions used for the electronic coordinates. In this case, it is customary to adopt periodic Born-von Karman boundary conditions that greatly simplify the classification of electronic states using the Bloch theorem. 

A dipole coupling of Bloch electrons with an electromagnetic field is commonly described using the velocity gauge. We, therefore, start with the following generic Hamiltonian of a periodic one-dimensional system (in the second quantization formalism), interacting with a set of long wavelength cavity modes~\cite{PhysRevB.101.205140,PhysRevResearch.2.033033,PhysRevLett.125.217402,PhysRevLett.125.257604} (throughout the paper, we set $\hbar=1$ and use Gaussian units),
\begin{equation}
\begin{gathered}
\hat{H}=\hat{H}_{el}+\hat{H}_{ph}=\text{\ensuremath{\sum_{k}\hat{\psi}_{k}^{\dagger}\mathbf{h}(k-\hat{A})\hat{\psi}_{k}}}\\
+\frac{1}{2}\sum_{\alpha}\left[\hat{\pi}_{\alpha}^{2}+\omega_{\alpha}^{2}\hat{q}_{\alpha}^{2}\right]\:,\label{H-velocity}
\end{gathered}
\end{equation}
where the momentum $k$ belongs to the Brillouin zone, $\hat{\psi}_{k}^{\dagger}=(\hat{c}_{1,k}^{\dagger},\dots \hat{c}_{M,k}^{\dagger})$
is an $M$-component Fermi operator, $\mathbf{h}(k)$ is an $M\times M$ matrix Hamiltonian,
$\hat{A}=\sum_{\alpha}\lambda_{\alpha}\hat{q}_{\alpha}$ is the spatially uniform vector potential (with the factor of $e/c$ included in it)
related to the canonical photon coordinate operator of the $\alpha$th cavity mode $\hat{q}_{\alpha}=(2\omega_{\alpha})^{-1/2}(\hat{a}_{\alpha}^{\dagger}+\hat{a}_{\alpha})$, 
expressed in terms of standard Bose creation ($\hat{a}_{\alpha}^{\dagger}$)  and annihilation ($\hat{a}_{\alpha}$) operators, with coupling constant $\lambda_{\alpha}=e\sqrt{4\pi}E_{\alpha}$, where $E_{\alpha}$ is the amplitude of the mode function at the system location, and $\hat{\pi}_{\alpha}$ is the canonical photon momentum operator. The mode function is related to the electric field amplitude by dividing the latter by $\sqrt{\omega_{\alpha}}$. 
For simplicity, we assume a one-dimensional electronic system, but this is not essential and the multi-dimensional generalization is straightforward. As we focus on 1D systems here, the modes are assumed to be polarized along the chain, and the index $\alpha$ labels the frequencies of the cavity. The field is quantized in the Coulomb gauge when the dynamical degrees of freedom correspond to the transverse components of the vector potential $\hat{A}$. In contrast, the scalar potential does not have independent dynamics as it is fixed by the charge distribution and included in the Coulomb interaction, which is not of concern in this work. In the dipole approximation, $\hat{A}$ is approximated by a 
constant vector, and we obtain the Hamiltonian above.

While the velocity-gauge electron-photon coupling in Eq.~\eqref{H-velocity} looks structurally simple,  using it in practice can be very nontrivial technically. In general, in crystals, the one-particle Hamiltonian $\mathbf{h}(k)$ in Eq.~\eqref{H-velocity} can be highly non-polynomial, as is the case, for example, within the tight-binding description of the band structure. This will make the standard field theoretical perturbative description of such systems enormously difficult beyond the simplest Gaussian approximation. The formal reason is that the expansion of a complicated function $\mathbf{h}(k-\hat{A})$ in powers of the vector potential $\hat{A}$ will generate bare multi-leg electron-photon vertices up to infinite order. This problem has much in common with difficulties in the description of the non-linear optical response of solids using the velocity gauge \cite{Aversa1995,Peres2017,Peres2018}. A possible way to circumvent this problem is to eliminate the vector potential from the electronic kinetic energy by transforming the Hamiltonian to the length gauge.

It seems indeed formally possible to gauge away the uniform vector potential $\hat{A}$ from the
electronic Hamiltonian by making the following unitary transformation (see  Appendix A for details):
\begin{equation}
e^{i\hat{X}\hat{A}}\hat{H}e^{-i\hat{X}\hat{A}} \label{UnitaryTransformation}\, ,
\end{equation}
where the `center-of-mass' position operator $\hat{X}$ is defined as
\begin{equation}
\hat{X}=i\sum_{k}\hat{\psi}_{k}^{\dagger}\partial_{k}\hat{\psi}_{k} \equiv \sum_{k,k'}\hat{\psi}_{k}^{\dagger}\mathbfcal{X}^{k,k'}\hat{\psi}_{k'}\:. \label{CO1}
\end{equation}
Throughout the paper, we use the convention that boldface symbols (e.g. $\mathbf{h}$) without any indices stand for single-body operators, which are represented by matrices in both, momentum and band space; if a boldface symbol has momentum (or band) index (e.g. $\mathbf{h}^{k,k^\prime}$ or $\mathbf{h}_{n,n'}$), it means that it remains a matrix in the momentum (or band) space; if the boldface symbol has a momentum dependence in the brackets, it just implies that the matrix is diagonal in the momentum space (i.e., for example,  $\mathbf{h}_{k,k^\prime}=\mathbf{h}(k)\delta_{k,k^\prime}$) or if the matrix is non-diagonal in the momentum space, then this notation implies that we take only the diagonal part. The corresponding second-quantized operators are denoted with hats.

Having established the notation convention, the canonical transformation to new variables,
\begin{align} \label{canonic-transform}
    \hat{q}_{\alpha}\text{\ensuremath{\mapsto\frac{1}{\omega_{\alpha}}\hat{P}_{\alpha},\qquad\hat{\pi}_{\alpha}\mapsto -\omega_{\alpha}\hat{Q}_{\alpha}}}
\end{align}
with $[\hat{Q}_{\alpha},\hat{P}_{\beta}]=i\delta_{\alpha,\beta}$,  brings the electron-photon Hamiltonian to the standard length-gauge form
\begin{equation} 
\begin{gathered}
\hat{H}=\sum_{k}\hat{\psi}_{k}^{\dagger}\mathbf{h}(k)\hat{\psi}_{k}\\
+\frac{1}{2}\sum_{\alpha}\left[\hat{P}_{\alpha}^{2}+\omega_{\alpha}^{2}\left(\hat{Q}_{\alpha}-\frac{\lambda_{\alpha}}{\omega_{\alpha}}\hat{X}\right)^{2}\right]\label{H-length}\:,
\end{gathered}
\end{equation}
also known as the Power-Zienau-Woolley Hamiltonian~\cite{Power1957,Woolley1971}. Physically, the photon momentum $\hat{P}_{\alpha}$ in the length gauge corresponds to the magnetic field, $\omega_{\alpha}\hat{Q}_{\alpha}$ is the electric displacement in the $\alpha$-mode, and $\lambda_{\alpha}\hat{X}$ is the polarization of the electronic system projected onto the $\alpha$-mode. Accordingly, the combination $\omega_{\alpha}\hat{Q}_{\alpha} - \lambda_{\alpha}\hat{X}$, entering the last term in Eq.~\eqref{H-length}, has the meaning of an electric field. 

The interaction part of the length-gauge Hamiltonian explicitly reads,
\begin{equation}
\begin{gathered}
    \hat{H}_{\text{int}}=\sum_{\alpha}\bigg[-\omega_{\alpha}\lambda_{\alpha}\hat{Q}_{\alpha}\hat{X}+\frac{\lambda_{\alpha}^2}{2}\hat{X}^2\bigg]\:.\label{H_int}
\end{gathered}
\end{equation}
The first term in the interaction Hamiltonian \eqref{H_int} describes the standard ``three-leg'' fermion-boson (two fermion- and one boson- operators) coupling between electrons and cavity photons. The second term accounts for a dipole self-energy which enters as an additional instantaneous electron-electron interaction \cite{Pellegrini2015PRL,Tokatly2018}.

An obvious advantage of the length-gauge representation, Eq.~\eqref{H-length}, over the velocity gauge, Eq.~\eqref{H-velocity}, is that the interaction Hamiltonian given by Eq.~\eqref{H_int} is a low-order polynomial in the Fermi and Bose fields, which has a structure very common in condensed matter physics. This dramatically simplifies the application of standard field theoretical methods to the description of this system. 

The structure of the diagram technique based on the length-gauge Hamiltonian is practically obvious from the form of its interaction part, Eq.~\eqref{H_int}. In fact, it can be  brought~\cite{Tokatly2018}  to the form identical to that for most standard fermion-boson Hamiltonians (for instance, for electron-phonon Hamiltonians). The basic elements of the diagram technique are the standard bare electron Green function and the following bare photon propagator,
\begin{equation}
    D_{\alpha}(t-t')=\omega_{\alpha}^2\braket{\hat{Q}_{\alpha}(t)\hat{Q}_{\alpha}(t')}+\delta(t-t')\:, \label{PhotonPropagatorTD}
\end{equation}
which combines the effect of two interaction terms in Eq.~\eqref{H_int}. Here, the first term $\sim \braket{\hat{Q}_{\alpha}(t)\hat{Q}_{\alpha}(t')}$ is the bare propagator of the photon canonical coordinate (the displacement propagator) while the second instantaneous term in Eq.~\eqref{PhotonPropagatorTD} comes from the second interaction term in Eq.~\eqref{H_int}. In total, Eq.~\eqref{PhotonPropagatorTD} describes the propagator of the cavity electric field, which mediates the physical effective electron-electron interaction induced by the coupling to cavity modes \cite{Tokatly2018}. In the Matsubara frequency representation, the bare photon propagator is the inverse Fourier transform of $D_{\alpha}(t-t')$ and it reads
\begin{equation}
    D_{\alpha}(i\omega_n)=\frac{\omega_\alpha^2}{(i\omega_n)^2-\omega_\alpha^2}+1=\frac{\omega_n^2}{\omega_n^2+\omega_{\alpha}^2}\:. \label{PhotonPropagatorTime}
\end{equation}
where $\omega_n=2\pi n/\beta$,  are bosonic Matsubara frequencies, with $n$ integer and $\beta=1/(k_BT)$ the inverse temperature ($k_B$ Boltzmann constant). The transformation from the real frequency representation for the bare photon propagator in the length gauge given by Eq.~(9) in Ref.~\cite{Tokatly2018} to the expression in Matsubara frequencies is done using the conventional substitution. 

\begin{figure}[!tb]
\includegraphics[width=0.12\textwidth]{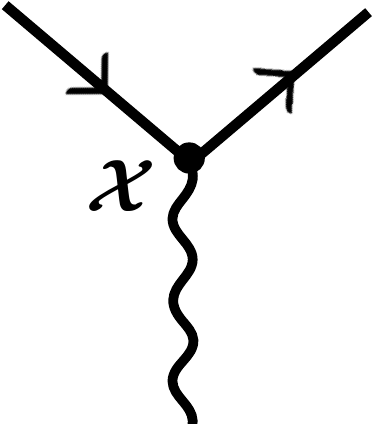}
\caption{Electron-photon interaction vertex in the length gauge, originating from the interaction Hamiltonian, given by Eq.~\eqref{H_int}. Here, $\mathbfcal{X}$ is a one-particle position operator. The solid lines correspond to electrons, while the wavy line accounts for photons. } \label{IntVertex}
\end{figure}

Within this formalism, there is only one type of three-leg electron-photon interaction vertices, shown in Fig.~\ref{IntVertex} and determined by the one-particle position operator $\mathbfcal{X}$,
as defined in Eq.~\eqref{CO1}. 
Apparently, the problem of properly defining the position operator in a periodic system is of critical importance for the diagram technique in the length gauge as it enters the interaction vertex. One can easily see that the operator $\hat{X}$ in Eq.~\eqref{CO1}, which generates the transformation from the velocity to the length gauge, is identical to the position operator introduced in the classic work by Bluont~\cite{Blount1962} (see also Ref.~\cite{Peres2018}). In the Bloch representation for an infinite periodic system, this operator takes the following form,
\begin{equation}
\mathcal{X}^{k,k^\prime}_{n,n^\prime}=i\delta_{n,n^\prime}\nabla_k\delta(k'-k)+\mathcal{A}_{n,n^\prime}(k)\delta(k'-k)\:,\label{CO_Bluont}
\end{equation}
where $\mathcal{A}_{n,n'}(k)=i\braket{u_{n,k}|\partial_ku_{n',k}}$.  A detailed discussion can also be found in Ref.~\cite{Parker2019}. Unfortunately, the position operator of Eq.~\eqref{CO_Bluont} is highly singular and not really well-defined mathematically. In fact, it requires from the beginning to be in the limit when the momentum space is a continuum and should always be treated in a distribution sense. This appears as the price to pay for the simple structure of the diagrammatic perturbation theory in the length gauge.   

In a related problem of nonlinear optical responses, the above technical difficulties are overcome by computing the response from the equation of motion for the density matrix. In this formalism, the position operator defined by Eq.~\eqref{CO_Bluont} appears only inside commutators, which eliminates all ambiguities related to its singular nature. This approach has been suggested in Ref.~\cite{Aversa1995} and used subsequently by many authors \cite{Peres2017,Peres2018,Parker2019}. Whether a similar reformulation is possible for the field-theoretical perturbation theory is an open question.

An alternative way to handle the problem is to redefine the position operator appearing in the electron-photon interaction vertex. We have seen that the position operator is a natural generator of the transformation from the velocity to the length gauge, that is, the transformation that gauges away the vector potential from the electronic 
Hamiltonian. The simplest way of avoiding singular distributions in the momentum space is to consider a large, but finite system with periodic boundary conditions over the system size $L$. However, by examining the unitary transformation \eqref{UnitaryTransformation} we observe that it is not consistent with the periodic boundary conditions for the electronic fields $\hat{\text{\ensuremath{\psi}}}(x)=\hat{\text{\ensuremath{\psi}}}(x+L)$.  This underlines a fundamental physical fact -- a uniform
vector potential can not be gauged away from a finite periodic system. The reason is that a vector potential
with a non-integer (in units $2\pi$) circulation produces a nontrivial
magnetic flux (modulo flux quantum) through the ring. It thus generates
a persistent current -- a physical effect that a gauge transformation can not eliminate. Only in the strict thermodynamic limit $L\to\infty$ the transformation defined by Eqs.~\eqref{UnitaryTransformation}-\eqref{CO1} works and the constant vector potential can be gauged away from the electronic Hamiltonian (i.e. the persistent current vanishes, typically as $1/L$). This can be viewed as another side of the statement that the position operator Eq.~\ref{CO_Bluont} requires a strictly infinite periodic system to be well defined. 

This observation suggests a new approach for the length gauge for a finite periodic system where we introduce a space-dependent $L$-periodic pure-gauge vector potential $\hat{A}(x)=\hat{A}(x+L)$. 
At the classical level of the electromagnetic field, such a pure-gauge potential does not cause any physical effect, and, indeed,  can be eliminated entirely from the problem by a unitary gauge transformation on the electronic Hamiltonian, due to the fact that the periodic boundary conditions of the wavefunctions remain satisfied. 

The generator of this transformation will then be identified as a regularized position operator. In this approach, the following limit, which we refer to as "continuous limit" should be taken: the sums over momenta are replaced by integrals accordingly, whereas the mode non-uniformity is eliminated (we will introduce below the parameter $q$ responsible for the deviation of the mode profile from perfectly uniform), preserving only the leading terms stemming from the fact that the parameter $q$ is $L$-dependent. We emphasize that $L=Na$ remains finite, but large and in any given order of the $1/N$ expansion (see below) we employ here. Replacing sums over momenta with integrals implies that the characteristic energy scales of interest are larger than the level spacing, caused by the finite size of the system.

However, as soon as we consider a quantized electromagnetic field,  $\hat{A}(x)$ cannot be gauged away and observable effects emerge from it in this quantum regime as will be discussed in the following.

\section{Regularization of the position operator.}

Below, we follow the program outlined at the end of the previous section to construct a regularized position operator in the electron-photon vertex. Again for the sake of simplicity, we consider a one-dimensional system of electrons coupled to the quantum cavity modes but now subjected
to periodic boundary conditions over a finite length $L$, which identify points $x$ and $x+L$.
In addition, we assume the presence of a periodic potential $V(x)=V(x+a)$,
such that the system contains in total $N$ unit cells, $L=Na$. The length of the system is assumed to be finite but large compared to the unit cell such that  $N\gg 1$.  The
second-quantized Hamiltonian now reads,
\begin{equation}
\begin{gathered}
\hat{H}=\int dx\:\bigg\{ \hat{\psi}^{\dagger}(x)\frac{1}{2m}\left[-i\nabla-\hat{A}(x)\right]^{2}\hat{\text{\ensuremath{\psi}}}(x)\\
+V(x)\hat{\psi}^{\dagger}(x)\hat{\text{\ensuremath{\psi}}}(x)\bigg\} +\hat{H}_{ph}\:,\label{H-periodic}
\end{gathered}
\end{equation}
where $\hat{A}(x)$ is a vector potential of the quantum electromagnetic field, and $H_{ph}$ is the photonic part of the Hamiltonian [the same as the last term in Eq.~\eqref{H-velocity}]. 

To formally define a finite-$L$ analog of the length gauge, we assume a pure-gauge vector potential $\hat{A}(x)$ can be represented as a gradient of an $L$-periodic (operator-valued) function,
\begin{equation}
\hat{A}(x)=\nabla\hat{\theta}(x)\:,\quad\text{with}\quad\hat{\theta}(x+L)=\hat{\theta}(x)\:.\label{A-condition}
\end{equation}

As a simple explicit realization, which reduces to a constant in the
limit $L\to\infty$, one can take the following space-periodic functions

\begin{equation}
\begin{gathered}
\hat{A}(x) =\sum_{\alpha}\lambda_{\alpha}(x)\hat{q}_{\alpha}=\sum_{\alpha}\nabla\theta_{\alpha}(x)\hat{q}_{\alpha}\\
=\sum_{\alpha}\nabla\theta_{\alpha}(x)\frac{1}{\sqrt{2\omega_{\alpha}}}(\hat{a}_{\alpha}^{\dagger}+\hat{a}_{\alpha}),\label{A-standing}
\end{gathered}
\end{equation}
where
\begin{equation}
\begin{gathered}
\lambda_{\alpha}(x)=\sqrt{2}\lambda_{\alpha}\cos(qx)\:,\\
\theta_{\alpha}(x)=\sqrt{2}\lambda_{\alpha}\frac{\sin(qx)}{q}\text{, }\ \ q=\frac{2\pi l}{L}
\:.
\end{gathered}
\end{equation}
Here, $l$ is an integer which, for our purpose, should be taken sufficiently small $l\ll N$ to mimic a long wavelength photon mode. For definiteness, we assume $l=1$  below. The factor $\sqrt{2}$ in the expressions for $\lambda_{\alpha}(x)$ and $\theta_{\alpha}(x)$ comes from the normalization of the mode function for the vector potential. The extraction of $\sqrt{2}$ allows us to keep the definition of the coupling constant $\lambda_{\alpha}$ the same as for the constant vector potential. In other words, independently of the mode functions we keep fixed, the space averaged field intensity $\sim\sqrt{\langle\lambda_{\alpha}^2(x)\rangle}$, or equivalently the space averaged photon propagator. The finite value of $q$, describing the non-uniformity of the mode profile, is not related to the mode frequency and should not be confused with the photon momentum. The results should not depend on the particular choice of the mode profile as long as the limit of a uniform mode is taken at the end.

If the condition (\ref{A-condition}) is satisfied, the vector potential
is eliminated from the electronic Hamiltonian by a unitary transformation,
\begin{align}
e^{-i\hat{S}}\hat{H}_{el}[-i\nabla-\hat{A}]e^{i\hat{S}} & =\hat{H}_{el}[-i\nabla],\label{periodic-transfom}\\
\hat{S}=\int dx\ \hat{\theta}(x)\hat{\psi}^{\dagger}(x)\hat{\psi}(x) & =\sum_{\alpha}\hat{q}_{\alpha}\int dx\ \theta_{\alpha}(x)\hat{\psi}^{\dagger}(x)\hat{\psi}(x).\nonumber
\end{align}
Using the Baker–Campbell–Hausdorff formula, one can show that the field operator $\hat{\psi}(x)$ maps to $e^{-i\hat{S}}\hat{\psi}(x)e^{i\hat{S}}=e^{i\hat{\theta}(x)}\hat{\psi}(x)$, thus preserving the boundary conditions. Since the operator $\hat{S}$ is linear in the photon coordinates,
the photon momentum operator is transformed as 
\begin{equation}
\begin{gathered}
e^{-i\hat{S}}\hat{\pi}_{\alpha}e^{i\hat{S}}=\hat{\pi}_{\alpha}+i[\hat{\pi}_{\alpha},\hat{S}]=\hat{\pi}_{\alpha}+\int dx\:\theta_{\alpha}(x)\hat{\psi}^{\dagger}(x)\hat{\psi}(x).
\end{gathered}
\end{equation}
After the canonical transformation 
\begin{align} 
    \hat{q}_{\alpha}\text{\ensuremath{\mapsto\frac{1}{\omega_{\alpha}}\hat{ P}_{\alpha},\qquad\hat{\pi}_{\alpha}\mapsto -\omega_{\alpha}\hat{Q}_{\alpha}}},
\end{align}
the total transformed Hamiltonian $\hat{\tilde{H}}=e^{-i\hat{S}}\hat{H}e^{i\hat{S}}$
takes the form,
\begin{equation}
\begin{gathered}
\hat{\tilde{H}}=\int dx\:\hat{\psi}^{\dagger}(x)\left\{ \frac{-\nabla^{2}}{2m}+V(x)\right\} \hat{\text{\ensuremath{\psi}}}(x)\\
+\frac{1}{2}\sum_{\alpha}\left[\hat{P}_{\alpha}^{2}+\omega_{\alpha}^2\left(\hat{Q}_{\alpha}-\frac{1}{\omega_{\alpha}}\hat{{\cal P}}_{\alpha}\right)^{2}\right].\label{tildeH-periodic}
\end{gathered}
\end{equation}
Here, we introduced an $\alpha$-component of the polarization,
\begin{equation}
\begin{gathered}
\hat{{\cal P}}_{\alpha}=\int dx\:\theta_{\alpha}(x)\hat{\psi}^{\dagger}(x)\hat{\psi}(x)\\
=\sqrt{2}\lambda_{\alpha}\int dx\:\frac{\sin(qx)}{q}\hat{\psi}^{\dagger}(x)\hat{\psi}(x)\equiv\lambda_{\alpha}\hat{X}_{q}\:.\label{polarization-periodic}
\end{gathered}
\end{equation}
The operator in the last equality can be identified as a center-of-mass position operator adapted for a periodic system.
Equations (\ref{tildeH-periodic})-(\ref{polarization-periodic})
provide a consistent formulation in terms of the length-gauge Hamiltonian,
which can now be written in any convenient representation.

In particular, we can rewrite the position operator in Bloch representation which is convenient to describe a periodic system. 
In the Bloch representation, the Fermi field operator is written as
\begin{equation}\label{FFO}
\begin{gathered}
    \hat{\psi}(x)=\sum_{n,k}\psi_{n,k}(x)\hat{c}_{n,k}\:,\\
    \psi_{n,k}(x)=\frac{1}{\sqrt{L}}e^{ikx}u_{n,k}(x),
\end{gathered}
\end{equation}
where $k=2\pi {m}/L$, with ${m}$ integer, takes discrete values in the Brillouin zone, and where $n$ is the band index, with the following normalization conditions,
\begin{equation}\label{Normalization}
\begin{gathered}
    \int_L dx\: \psi_{n',k'}^*(x)\psi_{n,k}(x)=\delta_{k,k'}\delta_{n,n'}\:, \\
    \sum_{n,k}\psi_{n,k}^*(x)\psi_{n,k}(x')=\delta(x-x')\, ,
\end{gathered}
\end{equation}
and the anticommutation relation of Fermi operators is given by 
$\{\hat{c}_{n,k},\hat{c}_{n',k'}^{\dagger}\}=\delta_{n,n'}\delta_{k,k'}$.

Using Eqs.~\eqref{polarization-periodic}-\eqref{Normalization}, we write the center-of-mass position operator in Bloch representation as

\begin{equation}
\begin{gathered}
\hat{X}_{q}=\sum_{\substack{n,n' \\ k,k'}}\frac{i}{\sqrt{2}q}\left[\delta_{k',k+q}-\delta_{k',k-q}\right]
\langle u_{n,k}|u_{n',k'}\rangle\hat{c}_{n,k}^{\dagger}\hat{c}_{n',k'}\:,\label{X-q-final}
\end{gathered}
\end{equation}
where  $\langle u_{n,k}|u_{n',k'}\rangle=\int_{\substack{\text{unit}\\ \text{cell}}} dx\, u_{n',k'}^*(x)u_{n,k}(x)$ (see Appendix B).
As by construction $q=2\pi/L$, the limit of an infinite system $L\to\infty$ (or constant vector potential) corresponds to $q\to 0$. 

The position operator defined above in Eq.~\eqref{X-q-final} is a many-particle operator, which is a sum of one-particle operators with matrix elements
\begin{equation}
    \mathcal{X}^{k,k'}_{n,n'}=\frac{i}{\sqrt{2}q}\left[\delta_{k',k+q}-\delta_{k',k-q}\right]\braket{u_{n,k}|u_{n',k'}}, \label{CO_OneParticle}
\end{equation}
where $k$ and $n$ are the first indices of the matrix, $k'$ and $n'$ are the second indices. Equation \eqref{CO_OneParticle} defines a regularized position operator entering the electron-photon interaction vertex. Thus, at all intermediate steps, we can assume a discretized momentum space and $q=2\pi/L$, while the limit of large $L$ is taken at the end of the calculations. 

\section{Diagram technique in the length gauge}
In this section, we introduce and illustrate the diagram technique in the length gauge using the position operator of Eq.~\eqref{X-q-final}. First, we discuss the approximation adopted in this paper to demonstrate applications of the formalism. Next, we show how to calculate the dressed photon propagator and the cavity-induced correction to the expectation value of an arbitrary operator. Finally, we illustrate the application of the proposed diagram technique calculating macroscopic polarization and charge imbalance for the Rice-Mele model embedded in a single-mode cavity.\\

\subsection{General formulation of the diagram technique in the length gauge}
\subsubsection{Random Phase Approximation}

In the present work, for all specific calculations, we adopt an approximation that can be viewed as an analog of the random phase approximation (RPA) in the theory of an electron gas. In the path integral formulation, it is equivalent to the Gaussian approximation for the description of electromagnetic fluctuations. We note that at the Gaussian level, the calculations are also feasible in the velocity gauge \cite{Dmytruk2021,Dmytruk2022}. However, using the proposed diagram technique in the length gauge, one can easily include any desirable contribution beyond the RPA. 

Diagrammatically, the RPA corresponds to the dressing of the photon propagator by bare empty polarization loops, neglecting vertex corrections, see Fig.~\ref{DressedPhProp}. After that, when computing observables, one performs the expansion in powers of the dressed photon propagator and keeps the lowest order contribution. Assuming that the electron-photon coupling strength is inversely proportional to the square root of the mode volume $\lambda\sim 1/\sqrt{V_{\text{mode}}}\sim 1/\sqrt{N}$ the above expansion generates the expansion in powers of $1/N$. 

For an electronic system embedded in a cavity, in the formal limit of $N \to \infty$, the RPA gives the exact~\cite{note1overN} result (though strong electron-electron interaction may affect the validity of this statement~\cite{Passetti2022}). This can be understood by analyzing the order of magnitude of the polarization diagrams in the RPA series, Fig.~\ref{DressedPhProp}. Each bubble without vertices in the limit of a large system is proportional to $N$ because, by going to this limit, we replace the summation over the wave vectors by integration as follows, 
\begin{align}
 \sum_k \to \frac{Na}{2\pi}\int_{-\frac{\pi}{a}}^{\frac{\pi}{a}} dk\:. \nonumber
\end{align}
Each vertex gives an interaction constant $\lambda$, which is proportional to $1/\sqrt{N}$ (due to the normalization of the electromagnetic cavity mode). Finally, in the RPA each bubble has two vertices.  As a result, all diagrams in the RPA series are of the order of unity -- the factors of $1/N$ from the vertices in the corners are compensated by the factors of $N$ from the summation over the electronic excitations in the bubbles. Any additional diagram on top of the RPA, e.~g. vertex insertions bring more vertices and thus more factors of $1/N$ per bubble, uncompensated by the summations over the electronic states.  Therefore, their contribution vanishes in the limit $N\to\infty$. A similar argumentation regarding the importance of the term $\sim N\lambda^2$ in the limit of infinite $N$ is given in Ref.~\cite{Lenk2022}. As the rescaled coupling constant $\lambda\sqrt{N}$ is not the small parameter of our perturbative expansion, it can be large, allowing one to consider the (ultra) strong coupling limit.

The significance of the contribution $~1/N$ was also discussed in a number of other works from a slightly different point of view \cite{Andolina2019, Dmytruk2021, Roche2022}. For instance, in Refs.~\cite{Dmytruk2021, Roche2022} it was pointed out that it is necessary to consider Gaussian fluctuations at $1/N$ order in addition to the mean-field Hamiltonian, which diagrammatically means keeping the terms of the lowest order in the dressed photon propagator.

Here we note that
we are interested in the lowest order correction to the ground state properties of the electronic system, which is $\mathcal{O}(1/N)$. This correction vanishes in the strict thermodynamic limit (in contrast to the photon counterpart). Thus, strictly speaking, we are studying the cavity effect on electronic systems of large but finite sizes. We note that the foregoing statement (vanishing cavity effect in the thermodynamic limit) may need modification when excited states of the electronic system are involved~\cite{Eckhardt2022}.

\subsubsection{Dressed photon propagator in the RPA}

\begin{figure}[!t]
\includegraphics[width=0.45\textwidth]{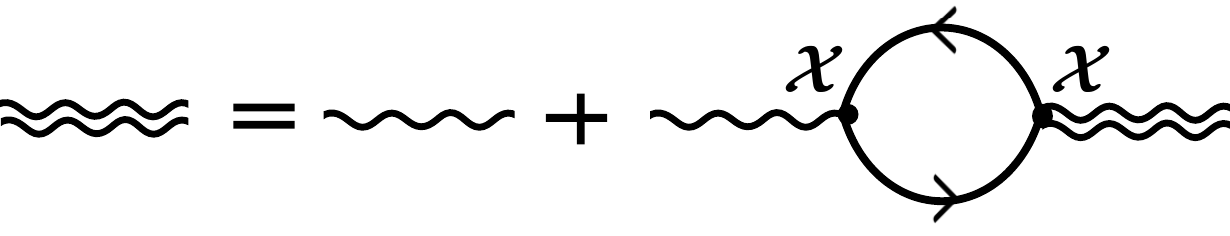}
\caption{Dyson equation for the dressed photon propagator in the RPA. Here, $\mathbfcal{X}$ is a one-particle position operator, given by Eq.~\eqref{CO_OneParticle}. The wavy line and double wavy line are bare photon propagators given by Eqs.~\eqref{PhotonPropagatorTD}, \eqref{PhotonPropagatorTime} and dressed photon propagator given by Eq.~\eqref{DressedPP}, respectively. The solid lines are bare electron Green functions, given in Eq.~\eqref{GFSpectrRepres}.}\label{DressedPhProp}
\end{figure}

Let us discuss how to perform the calculations in the RPA and start with the dressed photon propagator. The dressed photon propagator in the RPA is given by the infinite geometric series of diagrams, which can be written in the form of a Dyson equation represented in Fig.~\ref{DressedPhProp}.

From the Dyson equation, the dressed photon propagator in the RPA is found as
\begin{equation}
    D(i\omega_n)=\frac{D_0(i\omega_n)}{1+D_0(i\omega_n)\Pi(i\omega_n)}\:, \label{DressedPP}
\end{equation}
where $D_0(i\omega_n)=\omega_n^2/(\omega_n^2+\omega_0^2)$ is the bare photon propagator in the length gauge in Matsubara frequencies, Eq.~\eqref{PhotonPropagatorTime}, where $\omega_0$ is the frequency of the cavity mode (we assume that the parameters of the cavity are chosen such that only a single mode is relevant).
In the equation above,
$\Pi(i\omega_n)$  is the polarization operator in Matsubara frequencies given by
{\begin{equation}
\begin{gathered}
    \Pi(i\omega_n)=-\frac{g_0^2 }{Na\beta}\sum_{i\epsilon_m}\Tr\left[\mathbf{G}_0(i\epsilon_m+i\omega_n){\mathbfcal{X}}\mathbf{G}_0(i\epsilon_m)\mathbfcal{X}\right]\:,\label{PolOper}
\end{gathered}
\end{equation}
where $g_0=\lambda\sqrt{N a}$ and  we remind the reader that  $\beta^{-1}\sum_{i\epsilon_m}$ reduces to $\int d\epsilon_m/(2\pi)$ at  zero temperature. 
Here, Tr denotes the standard trace operation in momentum and band index space. 
The minus sign  
follows from the standard diagram rules for fermions  
in  Matsubara representation, where $\epsilon_m=(2m+1)\pi/\beta$, with $\beta=1/k_{B}T$ the inverse temperature and $m$ integer, are fermionic Matsubara frequencies.
We note that, as a result of the normalization convention for Bloch functions given by Eq.~\eqref{Normalization}, all $k$-summations in the present paper appear without an extra factor of $1/(Na)$.

Using the spectral representation for the bare electron Green function in the Bloch wave basis, 
\begin{equation}
   G^{k,k'}_{0n,n'}(i\epsilon_m)=\frac{\delta_{n,n'}\delta_{k,k'}}{i\epsilon_m-\epsilon_{n,k}}\label{GFSpectrRepres}\, ,
\end{equation}
where $\epsilon_{n,k}$ are the energy eigenvalues of the unperturbed electronic Hamiltonian,
and using
the matrix elements of the position operator given by Eq.~\eqref{CO_OneParticle}, the expression for the polarization operator $\Pi(i\omega_n)$ can be rewritten as
\begin{equation}
\begin{gathered}
    \Pi(i\omega_n)=\frac{g_0^2}{Na\beta}\sum_{i\epsilon_m}\sum_{\substack{n_1,n_2 \\ k_1,k_2}} \frac{1}{2q^2}
    \left[\delta_{k_2,k_1+q}-\delta_{k_2,k_1-q}\right]\\
    \times
    \braket{u_{n_1,k_1}|u_{n_2,k_2}}\frac{1}{i\epsilon_m-\epsilon_{n_2,k_2}}\left[\delta_{k_1,k_2+q}-\delta_{k_1,k_2-q}\right]\\
    \times\braket{u_{n_2,k_2}|u_{n_1,k_1}}\frac{1}{i(\epsilon_m+\omega_n)-\epsilon_{n_1,k_1}}\:. \label{PolOperCalc}  
\end{gathered}
\end{equation}

Considering that for arbitrary large, but finite $L$, $k$ is discrete, and $q=2\pi/L$ is finite as well, we obtain only two non-zero terms in Eq.~\eqref{PolOperCalc}. And after the summation over one of the wave vectors, we get the following expression
\begin{equation}
\begin{gathered}
    \Pi(i\omega_n)=-\frac{g_0^2}{Na\beta}\sum_{i\epsilon_m}\sum_{\substack{n_1,n_2,k}}\frac{1}{2q^2}|\braket{u_{n_1,k+q}|u_{n_2,k}}|^2
    \\
    \times\frac{1}{i\epsilon_m-\epsilon_{n_2,k}}\cdot
    \frac{1}{i(\epsilon_m+\omega_n)-\epsilon_{n_1,k+q}}
    +\left(q\to -q\right)\:.
\end{gathered}   
\end{equation}

After the summation over electron Matsubara frequencies $i\epsilon$ and taking the continuous limit of the multi-band electronic system, we obtain the following expression
\begin{equation}
\begin{gathered}
    \Pi(i\omega_n)=-g_0^2\sum_{n_1,n_2}\int_{-\frac{\pi}{a}}^{\frac{\pi}{a}}\frac{dk}{2\pi}\:|\braket{u_{n_1,k}|\partial_ku_{n_2,k}}|^2\\
    \times\frac{n_{\text{F}}(\epsilon_{n_2,k})-n_{\text{F}}(\epsilon_{n_1,k})}{\omega_n^2+(\epsilon_{n_2,k}-\epsilon_{n_1,k})^2}
    (\epsilon_{n_2,k}-\epsilon_{n_1,k})\:,\label{PolarOper}
\end{gathered}    
\end{equation}
where $n_{\text{F}}(\epsilon_{n,k})$ is the Fermi-Dirac distribution. Here, we used that $n_{\text{F}}(\epsilon_{n,k}\pm i\omega_n)=n_{\text{F}}(\epsilon_{n,k})$ as $\exp({\pm i \beta \omega_n})=1$.

From the expression for the dressed photon propagator [see Eq.~\eqref{DressedPP} with the polarization operator given by Eq.~\eqref{PolarOper}], we can straightforwardly obtain the photon spectral function. An example (photon spectral function for the SSH model~\cite{Su1979} embedded in a cavity) of applying the proposed diagram technique in the length gauge for its calculation can be found in Appendix C. It agrees with results obtained by Dmytruk and Schiro in Ref.~\cite{Dmytruk2022} using mean-field theory with the addition of Gaussian fluctuations in the velocity gauge.

\subsubsection{Correction to  one-particle observables}
Next, we calculate the correction to the expectation value of an arbitrary one-particle operator in the RPA due to the interaction of electrons with photons for a one-dimensional insulating electronic system embedded in a one-mode cavity.

In general, the average value of a one-particle operator
\begin{equation}
\hat{V}=\sum_{\substack{n,n'\\ k,k'}}\mathcal{V}_{n,n'}^{k,k'}\hat{c}^{\dagger}_{n,k}\hat{c}_{n',k'} \label{FermionicOperator}
\end{equation}
can be written in terms of the one-particle Green functions as (see Appendix D)
\begin{equation}  
\braket{\hat{V}}=\frac{1}{\beta}\sum_{i\epsilon_m}\text{Tr}\left[\mathbfcal{V}\mathbf{G}(i\epsilon_m)\right],
\end{equation}
where 
$G(i\epsilon_m)$ is a dressed electron Matsubara Green function.

\begin{figure}[t]
\includegraphics[width=0.18\textwidth]{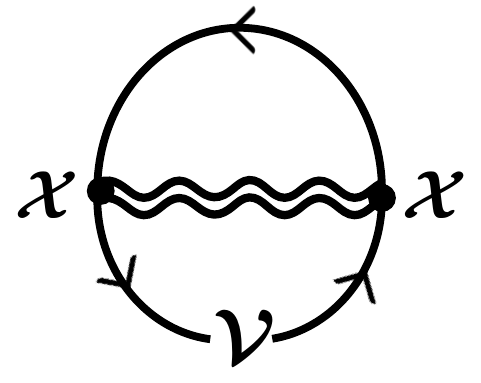}
\caption{Diagrammatic correction to the average value of an arbitrary operator $\hat{V}$ in the length gauge in the RPA. The analytical expression corresponding to the diagram is given in Eq.~\eqref{CorArbOper}. Here, $\mathbfcal{X}$ is the one-particle position operator given by Eq.~\eqref{CO_OneParticle}, $\mathbfcal{V}$ is defined in Eq.~\eqref{FermionicOperator}. Solid lines denote the bare electron Green functions, see Eq.~\eqref{GFSpectrRepres}. The double wavy line is a dressed photon propagator in the RPA given by Eq.~\eqref{DressedPP}. The Hartree diagram is identically zero because the photon propagator vanishes at zero frequency. This fact physically reflects the vanishing mean photon electric field in equilibrium.}\label{CorrAO}
\end{figure}

As we have discussed above in Sec.~IV~A1, within the length gauge formalism, the leading RPA correction to an observable is given by a diagram of the lowest order in the dressed photon propagator. The corresponding diagrammatic correction term for an observable related to an operator $\hat{V}$, which involves two vertices in terms of the position operator $\mathbfcal{X}$, is shown in Fig. \ref{CorrAO}. Analytically, the cavity-induced correction to the observable $\hat{V}$ takes the following form,
\begin{align}\label{CorArbOper}
    \delta V=-\frac{g_0^2}{Na\beta^2}\sum_{i\epsilon_m,i\omega_n}&\text{Tr}\Big\{D(i\omega_n) \mathbf{G}_0(i\epsilon_m)\\ &\times\mathbfcal{X}\mathbf{G}_0(i\epsilon_m+i\omega_n)\mathbfcal{X}\mathbf{G}_0(i\epsilon_m)\mathbfcal{V}\Big\}\:.\nonumber
\end{align}

This expression can be evaluated in the same way as the polarization operator above. Details of the calculation and the final result for an arbitrary operator diagonal in $k$-space can be found in Appendix E. These results will be used in the next subsection for calculating the cavity-induced charge imbalance in a  one-dimensional model.

\subsection{Application of the diagram technique - charge imbalance and polarization}
Now we illustrate the introduced diagram technique by applying it to the Rice-Mele model~\cite{RiceMele} placed in a single-mode cavity, where we evaluate two physical quantities of interest in the lowest order of perturbation expansion in the light-matter coupling: 
the charge imbalance and the macroscopic polarization, both in the dielectric (insulating) regime. 
\begin{figure}[b]
\includegraphics[width=0.45\textwidth]{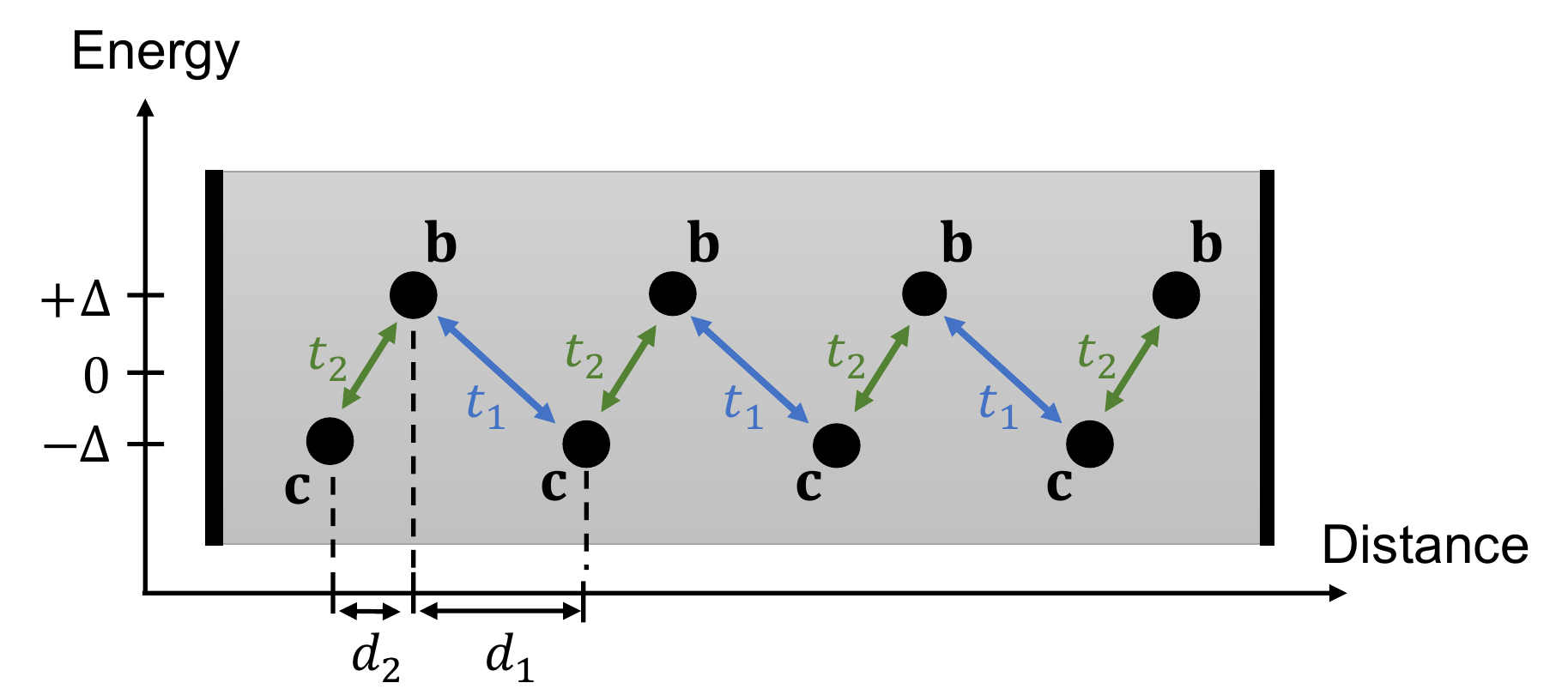}
\caption{Schematics of Rice-Mele model placed in a single-mode cavity. The right and left black bars indicate the cavity boundaries. Here, $2\Delta$ is the difference between the on-site energies of the two types of sites \textbf{b} and \textbf{c}. The distances from the site \textbf{b} to the nearest cite \textbf{c} on the right (left) is equal to $d_1$ ($d_2$). The corresponding hopping amplitudes are $t_1$ (blue) and $t_2$ (green).}
\label{fig:1}
\end{figure}
The considered system is schematically shown in Fig.~\ref{fig:1}, and its electronic part is described by the Hamiltonian,

\begin{equation}
\begin{gathered}
    \hat{H}_{\text{el}}=\sum_{m}\Bigg\{\Delta\left(\hat{b}^{\dagger}_m \hat{b}_m-\hat{c}^{\dagger}_m \hat{c}_m\right)\\
    + t_1\left(\hat{c}^{\dagger}_m \hat{b}_m+\hat{b}^{\dagger}_m \hat{c}_m \right)
    + t_2\left(\hat{c}^{\dagger}_m \hat{b}_{m+1}+\hat{b}^{\dagger}_{m+1}\hat{c}_m \right)\Bigg\}\:,
\end{gathered}
\end{equation}
where $2\Delta$ is the difference between the
on-site energies  
of the two types of sites ($\textbf{b}$ and $\textbf{c}$ sites),  $t_1$ and $t_2$ are the hopping amplitudes with  Fermi operators, $\hat{b}_m,\hat{c}_m$, corresponding to the two sites.
Introducing the Fourier transformations
\begin{equation}
\begin{gathered}
\hat{b}_m=N^{-1/2}\sum_k\hat{b}_ke^{ik(am-d_1)}\, ,\\
\hat{c}_m=N^{-1/2}\sum_k\hat{c}_ke^{ikam}\, , \label{FT}
\end{gathered}
\end{equation}
we obtain the Rice-Mele Hamiltonian in Fourier representation,
\begin{equation}
    \begin{gathered}
    \hat{H}_{\text{el}}=\sum_k\hat{\psi}^\dagger_k \mathbf{h}(k)\hat{\psi}_k\:,\\
    \mathbf{h}(k)=
        \begin{pmatrix}
        \Delta& t_1e^{ikd_1}+t_2e^{-ikd_2} \\
        t_1e^{-ikd_1}+t_2e^{ikd_2} & -\Delta
        \end{pmatrix}\:, \label{Rice-MeleHamiltonian}
    \end{gathered}
\end{equation}
where $k$ belongs to the Brillouin zone, $\hat{\psi}_k^{\dagger}=(\hat{b}_k^{\dagger}, \hat{c}_k^{\dagger})$, $d_1$ and $d_2$ are the distances between sites  
$\textbf{b}$ and $\textbf{c}$
 and sites $\textbf{c}$ and $\textbf{b}$, respectively, and $a=d_1+d_2$ is the lattice period of the system.
We note that in the absence of a cavity, we can make a unitary transformation and obtain an alternative form of the Rice-Mele Hamiltonian  
\begin{equation}
    \begin{gathered}
    \tilde{\mathbf{h}}(k)=
        \begin{pmatrix}
        \Delta& t_1+t_2e^{-ika} \\
        t_1+t_2e^{ika} & -\Delta
        \end{pmatrix}\:,
    \end{gathered}
\end{equation}
which is sometimes more convenient for calculations. However, in the presence of a cavity, such a unitary transformation would also change the photon propagator, so in our calculations, we use the Rice-Mele Hamiltonian in the form of Eq.~\eqref{Rice-MeleHamiltonian}.

The Rice-Mele model has a symmetric spectrum, i.e. $\epsilon_{+,k}=-\epsilon_{-,k}\equiv \epsilon_{k}$, which is found from the Schr\"odinger equation
\begin{equation}
    \mathbf{h}(k)\ket{u_{\pm, k}}=\epsilon_{\pm, k}\ket{u_{\pm, k}}\:,
\end{equation}
where $\ket{u_{\pm, k}}$ are Bloch eigenstates  for the upper ($+$) and lower ($-$) bands, respectively.

First, we calculate the correction to the charge imbalance. By definition, the charge imbalance is the charge difference between the sites $\textbf{b}$ and $\textbf{c}$,
\begin{equation}
\label{eq:charge_imb_def}
    \hat{\rho}=\frac{e}{N}\sum_m \left(\hat{b}^{\dagger}_m \hat{b}_m-\hat{c}^{\dagger}_m \hat{c}_m\right).
\end{equation}
In Fourier representation, the equation above becomes
\begin{equation}
    \hat{\rho}=\frac{e}{N}\sum_k \left(\hat{b}^{\dagger}_k \hat{b}_k-\hat{c}^{\dagger}_k \hat{c}_k\right). \label{ChargeImbalance}
\end{equation}
Obviously, the charge imbalance is an operator that is diagonal in $k$-space,  
so we can use the expression obtained for the cavity-induced correction to an arbitrary operator diagonal in $k$-space  (see Appendix F) and, as a result, the charge imbalance correction takes the form
\begin{equation}
\begin{gathered}
    \delta \rho_{\text{cav}}=\frac{eg_0^2}{N}\int_{-\frac{\pi}{a}}^{\frac{\pi}{a}} \frac{dk}{2\pi} \:[\varrho_{+,+}(k)-\varrho_{-,-}(k)]\,|\braket{u_{+,k}|\partial_k u_{-,k}}|^2\\
    \times\frac{1}{\beta}\sum_{i\omega_n} D(i\omega_n) \frac{\omega^2_n-4\epsilon_k^2}{(\omega^2_n+4\epsilon_k^2)^2}\\
        +\, 2\frac{eg_0^2}{N}\int_{-\frac{\pi}{a}} ^{\frac{\pi}{a}}\frac{dk}{2\pi}\:\text{Re}\big[\varrho_{+,-}(k)(\braket{u_{+,k}|\partial_k u_{+,k}}\\
        -\braket{u_{-,k}|\partial_k u_{-,k}})\braket{ u_{-,k}|\partial_k u_{+,k}}\big]
        \frac{1}{\beta}\sum_{i\omega_n}D(i\omega_n) \frac{1}{\omega^2_n+4\epsilon_k^2}\\
        +\, 2\frac{eg_0^2}{N}\int^{\frac{\pi}{a}}_{-\frac{\pi}{a}} \frac{dk}{2\pi}\:\text{Re}\left[\braket{u_{-,k}|\partial_k u_{+,k}}\epsilon_k\:\partial_k\left(\frac{\varrho_{+,-}(k)}{\epsilon_k}\right)\right]\\
        \times\frac{1}{\beta}\sum_{i\omega_n} D(i\omega_n)\frac{1}{\omega^2_n+4\epsilon_k^2}\:, \label{delta_rho}
\end{gathered}
\end{equation}
where $\varrho_{+,+}(k)$, $\varrho_{-,-}(k)$, and $\varrho_{+,-}(k)$ are matrix elements of the one-particle charge imbalance operator $\mathbf{\varrho}(k)$ in the Bloch wave basis, $D(i\omega_n)$ is the dressed photon propagator given by Eq.~\eqref{DressedPP} with the polarization operator Eq.~\eqref{PolarOper}.

To proceed, we have evaluated Eq.~\eqref{delta_rho} numerically and the result of the calculation is presented in Fig.~\ref{dSigma}.
In the present work, the energies are measured in units of $\omega_0$ (cavity frequency), and lengths are measured in the units of $a$ (period of the electronic system). In particular, Fig.~\ref{dSigma}  shows the dependence of the charge imbalance $\delta \rho_{\text{cav}}$ 
induced by the coupling to the cavity,
on $\Delta$, see Fig.~\ref{dSigma} (a),  
and on the hopping amplitude $t_2$ (with $t_1$ fixed), see Fig.~\ref{dSigma} (b), respectively.

\begin{figure}[t] 
\includegraphics[width=0.4\textwidth]{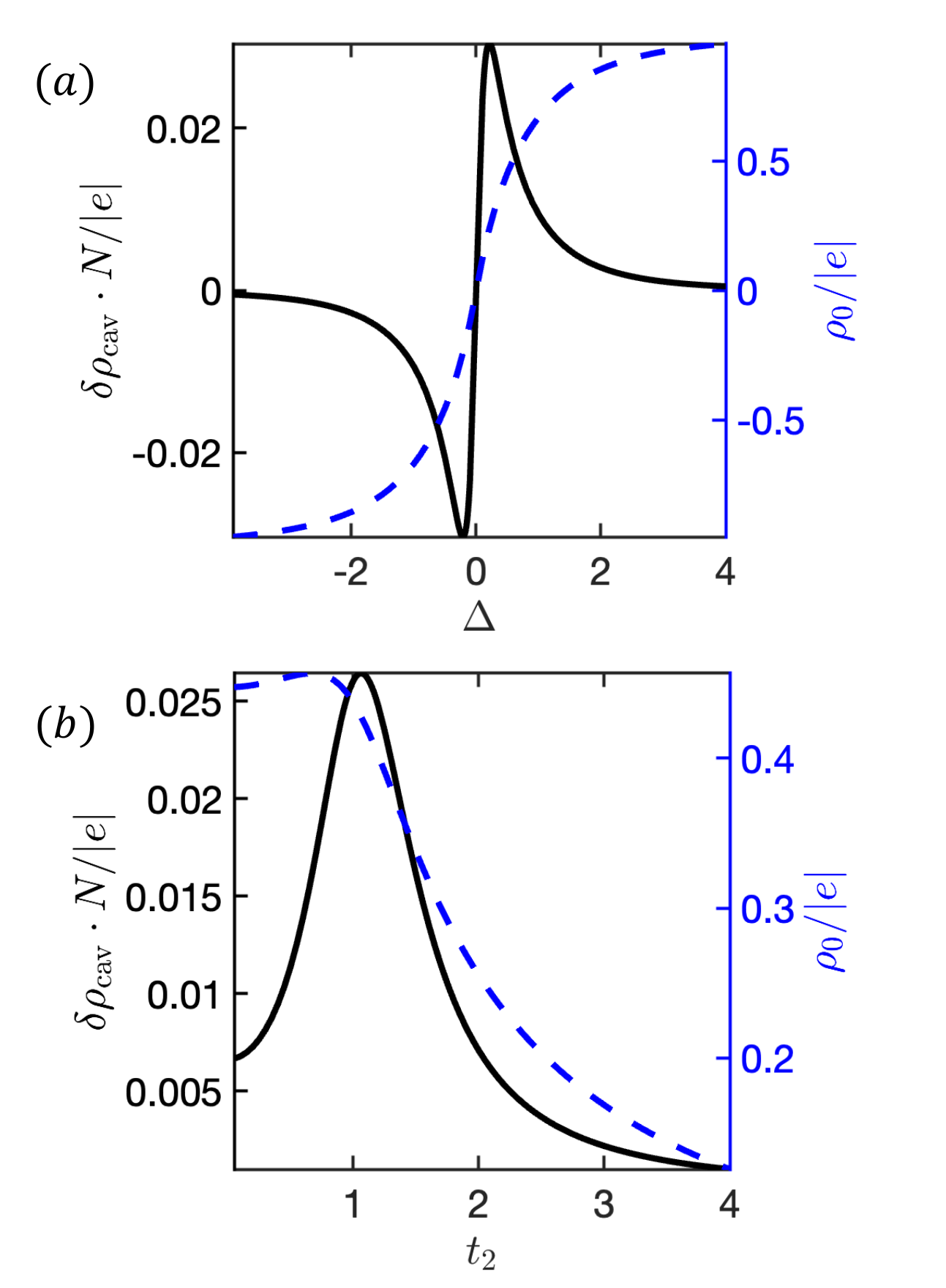}
\caption{The dependence of the cavity-induced charge imbalance correction $\delta \rho_{\text{cav}}$ given by Eq.~\eqref{delta_rho} (black solid curve) and the charge imbalance $\rho_0$ given by Eq.~\eqref{rho0} (blue dashed curve) for the Rice-Mele Hamiltonian coupled to a cavity: (a) on $\Delta$, where $2\Delta$ is the difference between the on-site energies
of the two types of sites, and (b) on the hopping amplitude $t_2$. Parameters: $t_1=1$, $g_0=0.8$, $d_1=d_2=0.5$, (a): $t_2=0.8$, (b): $\Delta=0.5$.  
}\label{dSigma}
\end{figure}
Suppose $\Delta$ takes large values in comparison to the other parameters of the system. In that case, the electronic system is nearly not affected 
by the presence of the electromagnetic field, so the correction to the charge imbalance due to the interaction of the electronic system with the electromagnetic field is negligibly small. From Fig.~\ref{dSigma} (a), it is seen that $\delta \rho_{\text{cav}} \to 0 $ in the limit $\Delta \to \pm \infty$. In the opposite limit ($\Delta = 0 $), there is no charge imbalance because the two types of sites have the same energies. In the same way as for large values of $\Delta$, the charge imbalance correction $\delta \rho_{\text{cav}}$ tends to zero in the limit $t_2 \to \infty$. The limit $t_2 \to 0$ corresponds to the case of $N$ separate dimers. 

For the unperturbed case, the non-monotonic dependence of charge imbalance
on the hopping amplitude, $t_2$  (see the blue dashed curve in Fig.~5) can be explained by analyzing  the analytical expression
\begin{equation}
    \begin{gathered}
    \rho_0=\frac{1}{\beta}\sum_{i\epsilon_m}\Tr\left[\mathbf{\varrho}(k)\:\mathbf{G}_0(k)\right]\\
    =-\frac{e}{N}\sum_k\frac{\Delta}{\sqrt{\Delta^2+t_1^2+t_2^2+2t_1t_2\cos{(ka)}}}\:. \label{rho0}
    \end{gathered} 
\end{equation}
Here, for the operators diagonal in momentum space, the trace operation $\Tr$ also includes the summation over a wavevector $k$ (in addition to the summation over band indices).
When $\cos{(ka)}>0$ in Eq.~\eqref{rho0}, the dependence of the integrand on $t_2$ is monotonically decreasing, while for $\cos{(ka)}<0$, the dependence is non-monotonic and has a maximum. Therefore, summing over all the states of the system, we still have a maximum in the dependence of the charge imbalance on 
$t_2$. The sign of the charge imbalance is positive, which follows from its definition Eq.~\ref{eq:charge_imb_def}: indeed, as $e <0$ and the energies of the \textbf{c}-sites are lower than the energies of the \textbf{b}-sites, one immediately concludes that more charges will be accumulated on the \textbf{c}-sites.

The correction to the charge imbalance, $\delta \rho_{\text{cav}}$, qualitatively behaves quite differently from the charge imbalance $\rho_0$ itself. Its dependence on the hopping amplitude $t_2$ is also non-monotonic but the peak is shifted and substantially more pronounced, which can serve as a means of distinguishing the correction from the bare contribution. 
The physical reason for the modification of the charge imbalance is the polaron effect~\cite{mahan1990many} (Sec. 7.1.2), which, in our case is induced by the vacuum fluctuations of the cavity mode, making it akin to the  Lamb shift. Namely, the electrons get dressed by the cavity photons and thus become heavier (which in terms of systems on a lattice implies that the hopping amplitudes decrease).

We would like to emphasize the difference between charge imbalance and polarization. In the modern (by now standard) theory~\cite{King-Smith1993,Resta1994,vanderbilt_2018}, the change in polarization
$\delta P$ in a periodic system is found by computing the current flowing
through the unit cell when some parameter $\xi(t)$ in the Hamiltonian
$\hat{H}(\xi)$ is changed adiabatically from some initial to a final
value, 
\begin{equation}
\delta P=\int_{t_{i}}^{t_{f}}J(t)dt\:. \label{DefPol}
\end{equation}
The current $J(t)$ can be calculated within linear response theory~\cite{Kotliar2013,bruus2004} to the perturbation of an adiabatic parameter $\xi(t)$ and, in this case, is described by the Kubo formula,
\begin{equation}
    J(t)=C_{\hat{J} \hat{H}'}^{(\xi)}(i\nu_m)\delta \xi(t)\:, \label{current}
\end{equation}
where $C_{\hat{J}\hat{H}'}^{(\xi)}(i\nu_m)$ is a zero-momentum component of the Fourier transform in Matsubara representation of the retarded correlation function,
\begin{align}
    C_{\hat{J}(r)\hat{H}'(r')}^{(\xi)}(t-t')=
    -i\theta(t-t')\braket{[\hat{J}(r,t),\hat{H}'_{\xi}(r',t')]}_{\xi}. \label{RetardedCF}
\end{align}
Here we assume the perturbation in the form $\delta \xi (t)=\delta \xi \exp{(-i\nu t +\eta t)}$ with $\nu$ being the real frequency related to the Matsubara frequencies $\nu_m$ and $\eta$ an infinitesimal positive number. In the above equation, $\hat{H}_{\xi}'$ is a functional derivative of the Hamiltonian with respect to the adiabatic parameter $\xi(t)$}, $\hat{J}$ is a current operator, and $\braket{...}_{\xi}$ denotes the averaging with respect to the instantaneous spectrum.
The correction to the correlation function $C_{\hat{J} \hat{H}'}^{(\xi)}(i\nu_m)$ in the RPA is shown in Fig. \ref{CorrelationFunction} as a sum of diagrams. It can be obtained from Fig. \ref{CorrAO} if we replace $\mathbfcal{V}$ to $\mathbfcal{J}$, where $\mathbfcal{J}$ is defined by $\hat{J}=\sum_{n,n'}\sum_{k}\mathcal{J}_{n,n'}(k)\hat{c}^{\dagger}_{n,k}\hat{c}_{n',k} $, and perturb all three Green functions in turns.

In Ref.~{\cite{Kotliar2013}}, Nourafkan and Kotliar formulated a diagrammatic approach for the calculation of the polarization correction using the Kubo formula written as
\begin{equation}
    J(t)=i\left(\frac{\partial}{\partial i \nu_m}C_{\hat{J} \hat{H}'}^{(\xi)}(i\nu_m)\right)\Bigg|_{\nu_m=0} \delta \dot \xi\:, \label{KuboFormula}
\end{equation}
which is obtained from Eq.~\eqref{current} in the limit of $\nu\to0$ (static distortion).
Furthermore, it was shown that the derivative of the polarization with respect to
an adiabatic parameter can be written as (see Eq.~(5) in Ref.~{\cite{Kotliar2013}})
\begin{equation}
\begin{gathered}
\frac{\partial P}{\partial\xi}=i\frac{e}{2N}\frac{1}{\beta}\sum_{i\epsilon_m}{\rm Tr}\Bigg\{ \mathbf{\Lambda}_{\mathcal{J}}(k)\frac{\partial \mathbf{G}(k)}{\partial i\epsilon_m}\mathbf{\Lambda}_{\xi}(k)\mathbf{G}(k)\\
-\mathbf{\Lambda}_{\xi}(k)\frac{\partial \mathbf{G}(k)}{\partial i\epsilon_m}\mathbf{\Lambda}_{\mathcal{J}}(k)\mathbf{G}(k)\Bigg\}\:,
\label{dP/dx-general}
\end{gathered}
\end{equation}
where the vertices $\mathbf{\Lambda}_{\mathcal{J},\xi}$ are related to the derivatives of $\mathbf{G}$ via the Ward
identities,
\begin{align}
\mathbf{\Lambda}_{\mathcal{J}}(k) & =-\frac{\partial \mathbf{G}^{-1}(k)}{\partial k}=-\frac{\partial \mathbf{G}_{0}^{-1}(k)}{\partial k}+\frac{\partial \mathbf{\Sigma}(k)}{\partial k}\label{Lambda-j}\:,\\
\mathbf{\Lambda}_{\xi}(k) & =-\frac{\partial \mathbf{G}^{-1}(k)}{\partial\xi}=-\frac{\partial \mathbf{G}_{0}^{-1}(k)}{\partial\xi}+\frac{\partial\mathbf{ \Sigma}(k)}{\partial\xi}\label{Lambda-x}\:,
\end{align}
where $\mathbf{\Sigma} (k)$ is the part diagonal in $k$  of the self-energy $\mathbf{\Sigma}$. We take only the diagonal part of the self-energy as we are interested in the current density integrated over space (which is nothing but its zero-momentum component)~\cite{Kotliar2013}.

\begin{figure}[t] 
\includegraphics[width=0.45\textwidth]{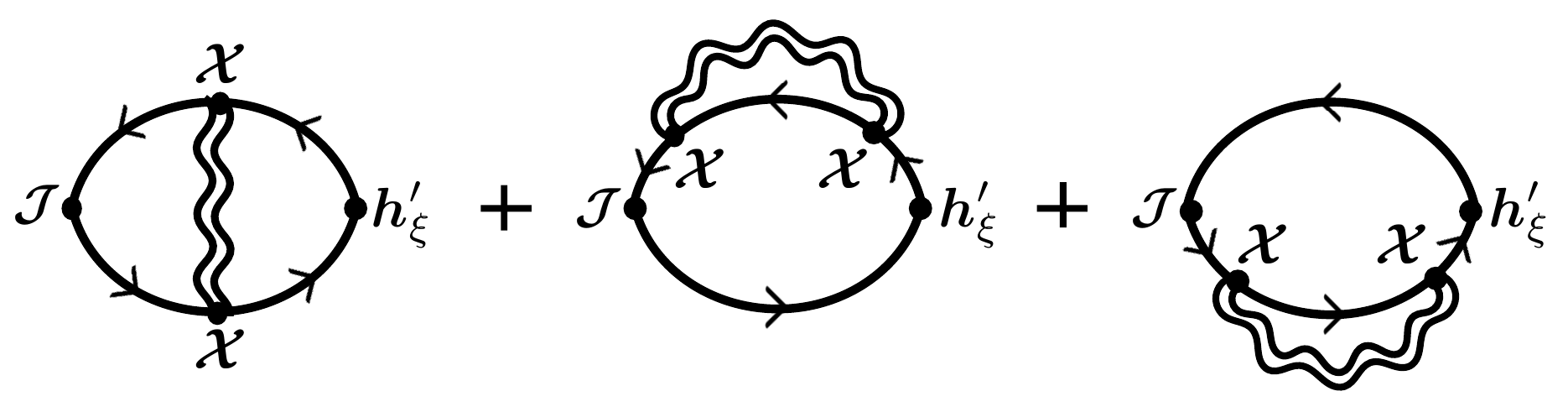}
\caption{Correction to the zero momentum component of the correlation function $C_{\hat{J}\hat{H}'}^{(\xi)}(i\nu_m)$ in the RPA, the real space representation of the correlation function is defined in Eq.~\eqref{RetardedCF}.  Here, $\mathbfcal{X}$ is a one-particle position operator given by Eq.~\eqref{CO_OneParticle}, $\mathbfcal{J}$ is a one-particle current operator, $\mathbf{h}'_{\xi}$ is a functional derivative with respect to the adiabatic parameter $\xi(t)$. The double wavy line is a dressed photon propagator given by Eq.~\eqref{DressedPP}. The solid line denotes a bare electron Green function.} \label{CorrelationFunction}
\end{figure}

For the lowest-order polarization correction ($\delta P =\delta P_0 + \delta P_{\text{cav}}+\mathbfcal{O}(1/N^2)$) Eqs.~\eqref{dP/dx-general} - \eqref{Lambda-x} reduce to (see Appendix G) 

\begin{equation}
    \frac{\partial P_{\text{cav}}}{\partial\xi}= i \frac{e}{2N} \frac{\partial}{\partial\xi}\frac{1}{\beta}\sum_{i\epsilon_m}{\rm Tr}\left\{ [\mathbf{G}_{0}(k),\partial_{k}\mathbf{G}_{0}(k)]\mathbf{\Sigma}(k)\right\}\:, \label{dP/dx-final}
\end{equation}
where 
\begin{equation}
[\mathbf{G}_{0}(k),\partial_{k}\mathbf{G}_{0}(k)]=\mathbf{G}_{0}(k)\left(\partial_{k}\mathbf{G}_{0}(k)\right)-\left(\partial_{k}\mathbf{G}_{0}(k)\right)\mathbf{G}_{0}(k)\:.\nonumber
\end{equation}

For the considered Rice-Mele model, we can choose $\Delta$ as an adiabatic parameter $\xi$. Then the correction to the polarization can be expressed as 
\begin{equation}
\begin{gathered}
    \delta P_{\text{cav}}=i\frac{e}{2N}\frac{1}{\beta}\sum_{i\epsilon_m}\text{Tr}\left\{[\mathbf{G}_{0}(k),\partial_{k}\mathbf{G}_{0}(k)]\mathbf{\Sigma}(k)\right\}(\xi=\Delta)\\
    -i\frac{e}{2N}\frac{1}{\beta}\sum_{i\epsilon_m}\text{Tr}\left\{[\mathbf{G}_{0}(k),\partial_{k}\mathbf{G}_{0}(k)]\mathbf{\Sigma}(k)\right\}(\xi=0)\:.
\end{gathered}
\end{equation}

Performing the calculation in the length gauge with the lowest-order self-energy 
\begin{equation}
    \mathbf{\Sigma}(i\epsilon_m)=(-1)\cdot\frac{g_0^2}{N\beta}\sum_{i\omega_n} D(i\omega_n)\mathbfcal{X}\mathbf{G}_0(i\epsilon_m+i\omega_n)\mathbfcal{X}\:,
\end{equation}
for a dielectric with two symmetric bands embedded in a cavity in the continuous limit, we obtain the following polarization correction
\begin{equation}
\begin{gathered}
    \delta P_{\text{cav}}= -i\frac{eg_0^2}{2N}\frac{1}{\beta}\sum_{i\omega_n}\int^{\frac{\pi}{a}}_{-\frac{\pi}{a}} \frac{dk}{2\pi} \:D(i\omega_n)\frac{\omega_n^2+12\epsilon_k^2}{(\omega_n^2+4\epsilon_k^2)^2}\\
    \times\bigg\{\braket{u_{+,k}|\partial_k u_{-,k}}\braket{u_{-,k}|\partial^2_k u_{+,k}}\\
    -\braket{u_{-,k}|\partial_k u_{+,k}}\braket{u_{+,k}|\partial^2_k u_{-,k}}
    +2|\braket{u_{-,k}|\partial_k u_{+,k}}|^2 \\   \times(\braket{u_{+,k}|\partial_ku_{+,k}}
    -\braket{u_{-,k}|\partial_k u_{-,k}})\bigg\}\:. \label{Polar}
\end{gathered}
\end{equation}
Details of the calculation can be found in Appendix H.

\begin{figure}[t] 
\includegraphics[width=0.4\textwidth]{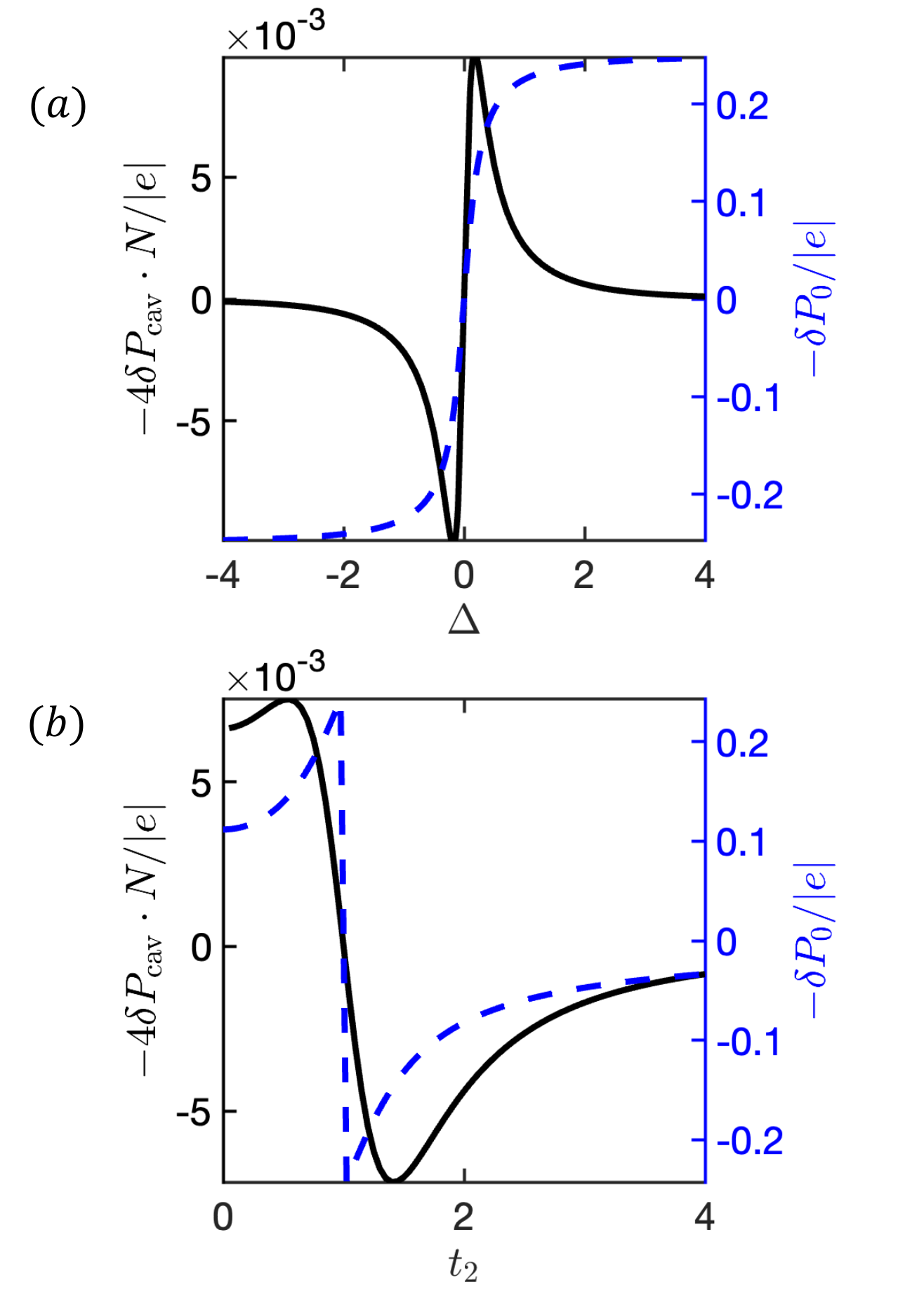}
\caption{The dependence of the cavity-induced polarization correction $\delta P_{\text{cav}}$ given by Eq.~\eqref{Polar} and the polarization $\delta P_0$ given by Eq.~\eqref{P0} for the Rice-Mele model embedded in a cavity: (a) on $\Delta$ where $2\Delta$ is the difference between the energy levels of the two types of sites and (b) on the hopping amplitude $t_2$. Parameters: $t_1=1$, $g_0=0.8$, $d_1=d_2=0.5$, (a): $t_2=0.8$, (b): $\Delta=0.5$.   
}\label{dP1}
\end{figure}

Figures \ref{dP1} (a,b) show the dependence of the correction to the polarization for the Rice-Mele model embedded in a cavity on $\Delta$ where $2\Delta$ is the difference between the 
on-site energies of the two types of sites and on the hopping amplitude $t_2$ while the hopping amplitude $t_1$ is fixed. We multiply the polarization correction by the coefficient $(-4)$ to satisfy the continuity equation in the limit of separate dimers ($t_2=0$). Its derivation for the present definitions of the charge imbalance and the current can be found in Appendix I. The sign of the polarization correction for the fixed $\Delta$ is determined by the direction of the current, which in turn is set by the ratio between the hopping amplitudes and changes when $t_1=t_2$. In the limits, $\Delta=0$ or $t_1=t_2$, the considered system has inversion symmetry,  which is consistent with the absence of polarization correction.

In the absence of a cavity Eqs.~\eqref{DefPol}--\eqref{KuboFormula} reduce to the Berry phase formula~\cite{vanderbilt_2018, Kotliar2013, King-Smith1993} 
\begin{equation}
\begin{gathered}
    \delta P_0=\frac{i e}{N}\sum_{k,n}\braket{u_{n,k}^{(\xi)}|\partial_k u_{n,k}^{(\xi)}}n_{\text{F}}(\epsilon_{n,k}^{(\xi)})(\xi=\Delta)\\
    -\frac{i e}{N}\sum_{k,n}\braket{u_{n,k}^{(\xi)}|\partial_k u_{n,k}^{(\xi)}}n_{\text{F}}(\epsilon_{n,k}^{(\xi)})(\xi=0)\:,\label{P0}
\end{gathered}
\end{equation}
where for the Rice-Mele model as an adiabatic parameter $\xi$ we can choose half of the difference between energy levels of the two types of sites and as initial and final states. 
We remind the reader that $\delta P_0$ is a change in polarization during the adiabatic process of varying $\Delta$, as the polarization itself is not unambiguous.

\section{Conclusion}
We developed a length-gauge formalism for the analysis of light-matter interaction in cavity-embedded electronic periodic systems. Despite being particularly useful for the diagrammatic analysis of light-matter correlations, the length-gauge formalism was hardly used in application to lattice models, which is mostly due to the long-standing problem of the ambiguities arising in the definition of a position operator in periodic systems. We have developed a method to eliminate this ambiguity which allows us to perform standard perturbative expansions for the calculation of observables. The crucial observation made in the present work is that it is important to keep the artificially introduced non-uniformity of the mode profile throughout the calculation, taking the limit in which the photon mode becomes uniform only at the very end of the calculation. We have applied the developed formalism to the problem of cavity-induced charge transfer and electric polarization in one-dimensional periodic systems. The formulation of the problem in the length gauge enables the use of conventional quantum field-theoretic methods to study the physics of ultrastrong coupling between light and matter in lattice models. We note that while in the thermodynamic limit, the leading contribution comes from RPA-type diagrams, for the mesoscopic systems, more complicated diagrams and associated processes will become essential. The proposed formalism allows one for performing an analysis of these processes using standard techniques. Our research enables new theoretical approaches to the description of ultrastrong light-matter coupling in crystals.

\begin{acknowledgments}
V.K.K. acknowledges the support from the Georg H. Endress Foundation. The work of I.V.T. was supported by Grupos Consolidados UPV/EHU del Gobierno Vasco (Grant IT1453-22) and by the grant PID2020-112811GB-I00 funded by MCIN/AEI/10.13039/501100011033. E.V.
acknowledges the support from NCCR SPIN. I.V.I acknowledges the support of “Basis” Foundation. We are grateful to
Dante M. Kennes for valuable discussions.
\end{acknowledgments}

\appendix
\numberwithin{figure}{section}
\section{Eliminating a constant vector potential from the electronic part of the Hamiltonian} 

The constant vector potential can be eliminated from the
electronic part of the Hamiltonian given by Eq.~\eqref{H-velocity} by the following unitary transformation introduced in Eq.~\eqref{UnitaryTransformation}:
\begin{equation}
 e^{-i\hat{X}\hat{A}}\hat{H}_{el}e^{i\hat{X}\hat{A}}=\ensuremath{\sum_{k}\hat{\psi}_{k}^{\dagger}\mathbf{h}(k)\hat{\psi}_{k}}\:,\label{Hel-length}
\end{equation}
where $\hat{X}$ is the polarization (dipole moment) operator,
\begin{equation}
\begin{gathered}
\hat{X}=i\sum_{k}\hat{\psi}_{k}^{\dagger}\partial_{k}\hat{\psi}_{k}=-i\sum_{k}(\partial_{k}\hat{\psi}_{k}^{\dagger})\hat{\psi}_{k}\\
=\frac{i}{2}\sum_{k}\left[\hat{\psi}_{k}^{\dagger}\partial_{k}\hat{\psi}_{k}-(\partial_{k}\hat{\psi}_{k}^{\dagger})\hat{\psi}_{k}\right].\label{X-def}
\end{gathered}
\end{equation}
To prove Eq.~(\ref{Hel-length}), we use the Hausdorff expansion,
\begin{equation}
\begin{gathered}
e^{-i\hat{X}\hat{A}}\hat{O}e^{i\hat{X}\hat{A}}=\hat{O}-\hat{A}[i\hat{X},\hat{O}]+\frac{\hat{A}^{2}}{2!}[i\hat{X},[i\hat{X},\hat{O}]]\\
-\frac{\hat{A}^{3}}{3!}[i\hat{X},[i\hat{X},[i\hat{X},\hat{O}]]]+\dots
\end{gathered}
\end{equation}
and apply it to the bilinear form $\hat{O}=\hat{c}_{n,k}^{\dagger}\hat{c}_{m,k}$. First
we evaluate the commutator with $\hat{c}$-operator,
\begin{equation}
\begin{gathered}
[i\hat{X},\hat{c}_{n,k}]=-\sum_{p,m}[\hat{c}_{m,p}^{\dagger}\partial_{p}\hat{c}_{m,p},\hat{c}_{n,k}]\\
=\sum_{p,m}\delta_{m,n}\delta_{p,k}\partial_{p}\hat{c}_{m,p}=\partial_{k}\hat{c}_{n,k}
\end{gathered}
\end{equation}
 and its Hermitian conjugate,
\begin{equation}
[i\hat{X},\hat{c}_{n,k}^{\dagger}]=\partial_{k}\hat{c}_{n,k}^{\dagger}\:.
\end{equation}
 By combining these results we find,
\begin{equation}
\begin{gathered}
[i\hat{X},\hat{c}_{n,k}^{\dagger}\hat{c}_{m,k}]=\hat{c}_{n,k}^{\dagger}[i\hat{X},\hat{c}_{m,k}]+[i\hat{X},\hat{c}_{n,k}^{\dagger}]\hat{c}_{m,k}\\
=\partial_{k}(\hat{c}_{n,k}^{\dagger}\hat{c}_{m,k})
\end{gathered}
\end{equation}
 and finally,
\begin{equation}
\begin{gathered}
e^{-i\hat{X}\hat{A}}\hat{c}_{n,k}^{\dagger}\hat{c}_{m,k}e^{i\hat{X}\hat{A}}  =\hat{c}_{n,k}^{\dagger}\hat{c}_{m,k}-\hat{A}[i\hat{X},\hat{c}_{n,k}^{\dagger}\hat{c}_{m,k}]\\
+\frac{\hat{A}^{2}}{2!}[i\hat{X},[i\hat{X},\hat{c}_{n,k}^{\dagger}\hat{c}_{m,k}]]+\dots\\
= \hat{c}_{n,k}^{\dagger}\hat{c}_{m,k}-\hat{A}\partial_{k}(\hat{c}_{n,k}^{\dagger}\hat{c}_{m,k})+\frac{\hat{A}^{2}\partial_{k}^{2}}{2!}(\hat{c}_{n,k}^{\dagger}\hat{c}_{m,k})+\dots\\
=e^{-\hat{A}\partial_{k}}\hat{c}_{n,k}^{\dagger}\hat{c}_{m,k}\:.
\end{gathered}
\end{equation}
Thus, the introduced operator $\hat{X}$ indeed acts on the one-particle
density-matrix operator as a generator of a shift in the momentum
space. Using this property, we can perform the transformation of the
electronic Hamiltonian, 
\begin{equation}
\begin{gathered}
e^{-i\hat{X}\hat{A}}\hat{H}_{el}e^{i\hat{X}\hat{A}}  =\sum_{n,m,k}h_{n,m}(k-\hat{A})e^{-\hat{A}\partial_{k}}\hat{c}_{n,k}^{\dagger}\hat{c}_{m,k}\\
=\ensuremath{\sum_{k}\hat{\psi}_{k}^{\dagger}\left[e^{\hat{A}\partial_{k}}\mathbf{h}(k-\hat{A})\right]\hat{\psi}_{k}}=\ensuremath{\sum_{k}\hat{\psi}_{k}^{\dagger}\mathbf{h}(k)\hat{\psi}_{k}}\:,
\end{gathered}
\end{equation}
which proves the result announced in Eq.~(\ref{Hel-length}) and, therefore, gives the electronic part of the electron-photon Hamiltoian in the length gauge [see Eq.~\eqref{H-length}].

\section{Position operator in Bloch representation}
The position operator for a periodic system with a single-mode vector potential is written as [see Eq.~\eqref{polarization-periodic} in the main text]
\begin{equation}
\begin{gathered}
\hat{X}_{q}=\sqrt{2}\int dx\:\frac{\sin(qx)}{q}\hat{\psi}^{\dagger}(x)\hat{\psi}(x)\:.
\end{gathered}
\end{equation}
In the Bloch representation, we get,
\begin{equation}
\hat{X}_{q}=\sum_{n,n',k,k'}\mathcal{X}^{k,k'}_{n,n'}\hat{c}_{n,k}^{\dagger}\hat{c}_{n',k'}\:,
\end{equation}
where
\begin{equation}
\begin{gathered}
\mathcal{X}^{k,k'}_{n,n'} =\sqrt{2}\int dx\:\frac{\sin(qx)}{q}\psi_{n,k}^{*}(x)\psi_{n',k'}(x)\\
=\frac{\sqrt{2}}{N}\int dx\:\frac{\sin(qx)}{q}e^{i(k'-k)x}u_{n,k}^{*}(x)u_{n',k'}(x)\\
 =\frac{i}{2q}\frac{\sqrt{2}}{N}\sum_{R}\int_{\substack{\text{unit}\\ \text{cell}}}dx\left[e^{i(k'-k-q)(R+x)}
 -e^{i(k'-k+q)(R+x)}\right]\\
 \times u_{n,k}^{*}(x)u_{n',k'}(x)
 =\frac{i}{\sqrt{2}q}\left[\delta_{k',k+q}-\delta_{k',k-q}\right]\langle u_{n,k}|u_{n',k'}\rangle\:.
\end{gathered}
\end{equation}
So, in the Bloch wave basis, we obtain the following expression for the position operator
\begin{equation}
\hat{X}_{q}=\sum_{\substack{n,n'\\ k,k'}}\frac{i}{\sqrt{2}q}\left[\delta_{k',k+q}-\delta_{k',k-q}\right]\langle u_{n,k}|u_{n',k'}\rangle\hat{c}_{n,k}^{\dagger}\hat{c}_{n',k'}\:,
\end{equation}
with $q=2\pi/L$, as it was announced in Eq.~\eqref{X-q-final}.

\section{Spectral function for SSH-model in a cavity as an example of applying the length-gauge formalism}

The photon spectral function is found as  
\begin{equation} 
    A(\omega)=-\frac{1}{\pi}\text{Im}\left[D(\omega)\right]\:.\label{SF1}
\end{equation}
At zero temperature, the photon propagator is written as
\begin{equation}
    D(\omega)=D^{(\text{R})}(\omega)\theta(\omega)+D^{(\text{A})}(\omega)\theta(-\omega)\:,
\end{equation}
where the retarded photon Green function $D^{(\text{R})}(\omega)$ and the advanced photon Green function $D^{(\text{A})}(\omega)$ can be obtained from the Matsubara Green function $D(i\omega_n)$ using the conventional substitution
\begin{equation}
\begin{gathered}
    D^{(\text{R})}(\omega)=D(i\omega_n\to \omega +i\delta)\\
    D^{(\text{A})}(\omega)=D(i\omega_n\to \omega -i\delta) \label{SF3}
\end{gathered}
\end{equation}
with $\delta>0$.\\

\begin{figure}[t]
\includegraphics[width=0.45\textwidth]{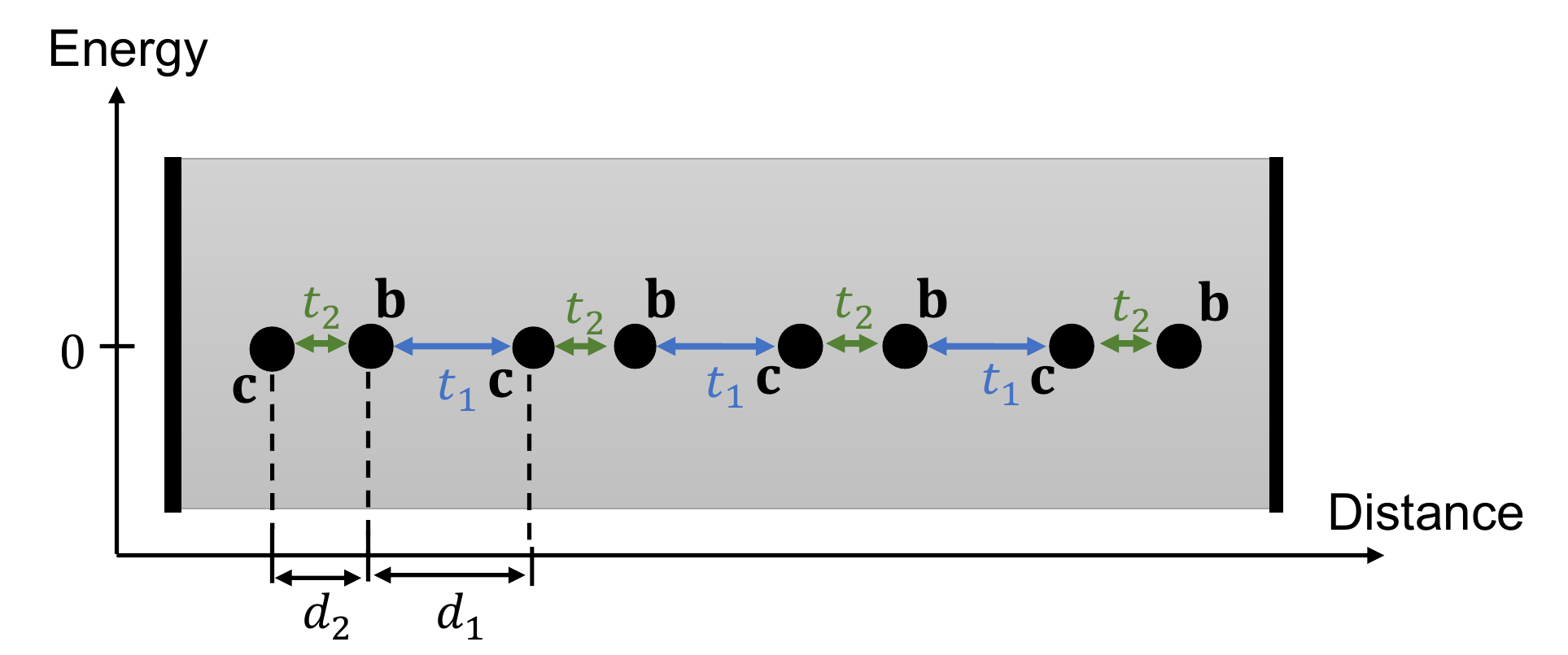}
\caption{Schematics of SSH model placed in a single-mode cavity. Black lines denote the cavity boundaries. The distances from the site \textbf{b} to the nearest cite \textbf{c} on the right (left) is equal to $d_1$ ($d_2$). The corresponding hopping amplitudes are $t_1$ (blue) and $t_2$ (green). In contrast to the Rice-Mele model shown in Fig.~\ref{fig:1}, there is no energy difference between the two types of sites.} \label{SSH_model}
\end{figure}

\begin{figure*}[t]
\includegraphics[width=0.95\textwidth]{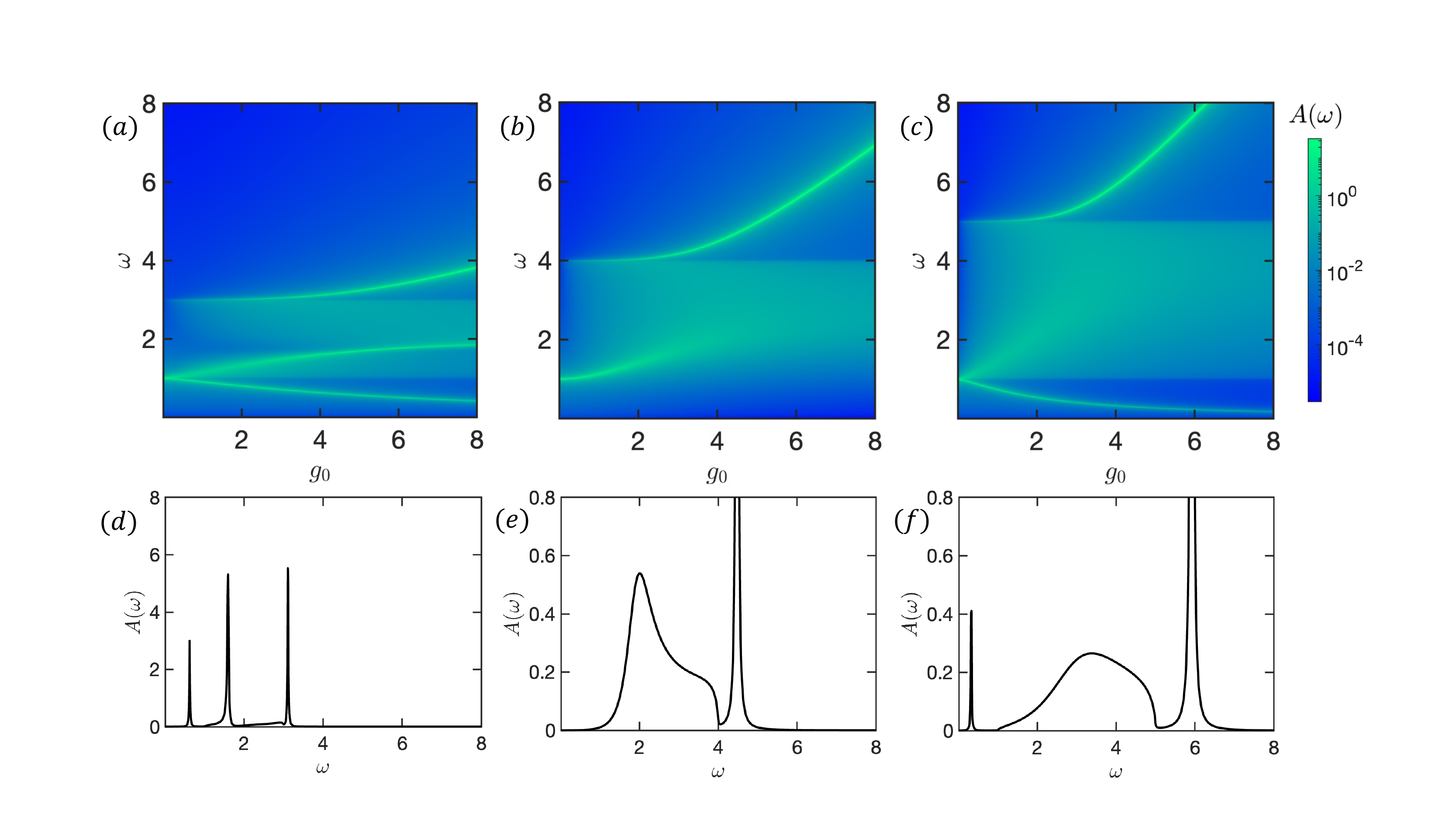}
\caption{Photon spectral density $A(\omega)$ for the SSH model embedded in a cavity [see Fig.~\ref{SSH_model}] as a function the coupling constant $g_0$ and frequency $\omega$ [(a)--(c)] and its profile for the the fixed coupling constant $g_0=4$ as a function of $\omega$ [(d)--(f)]. One of the hopping amplitudes is kept the same for all plots [$t_1=1$] while the second one differs [(a),(d): $t_2=-0.5$; (b),(e): $t_2=-1$; (c),(f): $t_2=-1.5$]. Other parameters are chosen as $d_1=1$, $d_2=0$, $\delta=0.01.$} \label{SpectralFunction}
\end{figure*}

As an example of applying our theory, we calculated the photon spectral density function for the SSH model~\cite{Su1979} embedded in a cavity which is schematically shown in Fig.~\ref{SSH_model}. The Hamiltonian of the SSH model reads as
\begin{equation}
    \begin{gathered}
    \hat{H}_{\text{el}}=\sum_k\hat{\psi}^\dagger_k \mathbf{h}(k)\hat{\psi}_k\:,\\
    \mathbf{h}(k)=
        \begin{pmatrix}
        0& t_1e^{ikd_1}+t_2e^{-ikd_2} \\
        t_1e^{-ikd_1}+t_2e^{ikd_2} & 0
        \end{pmatrix}\:, \label{SSH-Hamiltonian}
    \end{gathered}
\end{equation}
where $k$ belongs to the Brillouin zone, $\hat{\psi}_k^{\dagger}=(\hat{b}_k^{\dagger}, \hat{c}_k^{\dagger})$, $d_1$ and $d_2$ are the distances between sites inside the two-site unit cell, $t_1$ and $t_2$ are the corresponding hopping amplitudes. The SSH Hamiltonian represents a limiting case of the Rice-Mele Hamiltonian considered in the main text [see Eq.\eqref{Rice-MeleHamiltonian}] when there is no energy difference between the two types of sites.

The result of the calculation is presented in Fig.~\ref{SpectralFunction}, where we used Eqs.~\eqref{SF1} -- \eqref{SF3} and the length-gauge formalism in the RPA developed in the main text [see Eqs.~\eqref{DressedPP},\eqref{PolarOper}]. We chose the parameters for Fig.~\ref{SpectralFunction} in a similar way as in Ref.~\cite{Dmytruk2022}, where the same system was studied using mean-field theory with the addition of Gaussian fluctuations in the velocity gauge. The obtained result is in agreement with Fig.~5 in Ref.~\cite{Dmytruk2022}.

\section{Averaging of a one-particle operator}
Let us find the average value of a one-particle operator 
\begin{equation}
\hat{V}=\sum_{n,n'}\sum_{k,k'}
\mathcal{V}^{k,k'}_{n,n'}\hat{\psi}^{\dagger}_{n,k}\hat{\psi}_{n',k'}\:.
\end{equation}
The Fourier component of an annihilation operator is written as 
\begin{equation}
    \hat{\psi}_{n,k}=\frac{1}{\sqrt{L}}\int dr\ \hat{\psi}_n(x)e^{-ikx}\:.\label{FourierAO}
\end{equation}
Then using the Eq. \eqref{FourierAO}, one can calculate the average value of a one-particle operator $\hat{V}$ in the following way:

\onecolumngrid

\begin{align}
\begin{gathered} \bra{GS}\hat{V}\ket{GS}=\sum_{n,n'}\sum_{k,k'}\bra{GS}\mathcal{V}^{k,k'}_{n,n'}\hat{\psi}^{\dagger}_{n,k}\hat{\psi}_{n',k'}\ket{GS}=\frac{1}{L}\sum_{n,n'}\sum_{k,k'}\mathcal{V}^{k,k'}_{n,n'}\iint dr dr' \braket{GS|\hat{\psi}_n^{\dagger}(x)\hat{\psi}_{n'}(x')|GS}e^{ikx}e^{-ik'x'}\\
    =\frac{1}{L}\sum_{n,n'}\sum_{k,k'}\mathcal{V}^{k,k'}_{n,n'}\iint dx dx' \braket{GS|\hat{\psi}^{\dagger}_n(x,\delta)\hat{\psi}_{n'}(x',0)|GS}e^{ikx}e^{-ik'x'}\big|_{\delta\to+0}\\
    =-\frac{i}{i L}\sum_{n,n'}\sum_{k,k'}\mathcal{V}^{k,k'}_{n,n'}\iint dx dx'\braket{GS|\text{T}\left\{\hat{\psi}_{n'}(x',0)\hat{\psi}_n^{\dagger}(x,\delta)\right\}|GS}e^{ikx}e^{-ik'x'}\big|_{\delta\to+0}\\
    =-\frac{i}{L}\sum_{n,n'}\sum_{k,k'}\mathcal{V}^{k,k'}_{n,n'}\iint dx dx'\:G_{n',n}(x'-x,-\delta)e^{ikx}e^{-ik'x'}\big|_{\delta\to+0}\\
    =-\frac{i}{L}\sum_{n,n'}\sum_{k,k'}\mathcal{V}^{k,k'}_{n,n'}\iint d(x'-x) dx'\:G_{n',n}(x'-x,-\delta)e^{-ik(x'-x)}e^{i(k-k')x'}\big|_{\delta\to+0}\\
    =-i\sum_{n,n'}\sum_{k,k'}\mathcal{V}^{k,k'}_{n,n'}\delta_{k,k'}\int d(x'-x)\: G_{n',n}(x'-x,-\delta)e^{-ik(x'-x)}\big|_{\delta\to+0}=-i\sum_{n,n'}\sum_{k,k'}\mathcal{V}^{k,k'}_{n,n'}\delta_{k,k'}G_{n',n}(k,-\delta)\big|_{\delta\to+0}\\
    =-i\sum_{n,n'}\sum_{k,k'}\int\frac{d\epsilon}{2\pi}\:\mathcal{V}^{k,k'}_{n,n'}\delta_{k,k'}G_{n',n}(k,\epsilon)e^{-i\epsilon(-\delta)}\big|_{\delta\to+0}=-i\int\frac{d\epsilon}{2\pi}\:\text{Tr}[\mathbfcal{V}\mathbf{G}(\epsilon)]=\int\frac{d\epsilon_m}{2\pi}\:\text{Tr}[\mathbfcal{V}\mathbf{G}(i\epsilon_m)]\:,
\end{gathered}
\end{align}
\twocolumngrid
where $\epsilon_m$ is a fermionic Matsubara frequency.
\section{Correction to the average value of an arbitrary operator diagonal in $k$-space for a dielectric embedded in a cavity}
In this section, we calculate the cavity-induced correction to the average value of an arbitrary operator $\hat{V}$ given by Eq.~\eqref{CorArbOper} for the case when the operator $\hat{V}$ is diagonal in $k$-space. Using the spectral representation for the electron Green function in the Bloch wave basis and matrix elements of the one-particle coordinate operator, we rewrite Eq.~\eqref{CorArbOper} for an arbitrary operator $\hat{V}$ diagonal in $k$-space as 

\begin{align}
\begin{gathered} \label{ArbOperCorCalc}
    \delta V=\frac{g_0^2}{Na\beta^2}\sum_{i\epsilon_m,i\omega_n}\sum_{n_1,n_2,n_3}\sum_{k_1,k_2}\frac{D(i\omega_n)}{2q^2}\frac{1}{i\epsilon_m-\epsilon_{n_1,k_1}}\\
    \times\left[\delta_{k_2,k_1+q}-\delta_{k_2,k_1-q}\right]\braket{u_{n_1,k_1}|u_{n_2,k_2}}\\
   \times\frac{1}{i(\epsilon_m+\omega_n)-\epsilon_{n_2,k_2}}
   \left[\delta_{k_1,k_2+q}-\delta_{k_1,k_2-q}\right] \times \nonumber
\end{gathered} \\ 
 \begin{gathered}     \times\braket{u_{n_2,k_2}|u_{n_3,k_1}}
 \frac{1}{i\epsilon_m-\epsilon_{n_3,k_1}}\mathcal{V}_{n_3,n_1}(k_1)\:. 
\end{gathered}    
\end{align}

Taking into account that at this moment $q$ is finite, we obtain only two non-zero terms in Eq.~\eqref{ArbOperCorCalc}. And after the summation over one of the wave vectors, we get the following expression

\begin{equation}
    \begin{gathered}
        \delta V=-\frac{g_0^2}{Na\beta^2}\sum_{i\epsilon_m,i\omega_n}\sum_{n_1,n_2,n_3}\sum_k\frac{D(i\omega_n)}{2q^2}\frac{1}{i\epsilon_m-\epsilon_{n_1,k}}\\
        \times\frac{1}{i(\epsilon_m+\omega_n)-\epsilon_{n_2,k+q}}
        \frac{1}{i\epsilon_m-\epsilon_{n_3,k}}       \braket{u_{n_1,k}|u_{n_2,k+q}}\\
        \times\braket{u_{n_2,k+q}|u_{n_3,k}}\mathcal{V}_{n_3,n_1}(k)
        +\left(q\to -q\right)\:.
    \end{gathered}
\end{equation}
After that, we perform the summation over the electron Matsubara frequency $i\epsilon_m$, consider the continuous limit of the dielectric system and make the Taylor expansion up to the second order in $q$ (higher orders of the expansion will automatically go to zero if we take the limit $q \to 0$). As a result, we obtain the following expression:

\begin{equation}
    \delta V=\delta V_{\text{diag}}+\delta V_{\text{non-diag}}\:,
\end{equation}

\begin{equation}
\begin{gathered}
    \delta V_{\text{diag}}=\frac{g_0^2}{\beta}\sum_{i\omega_n}\int_{-\frac{\pi}{a}}^{\frac{\pi}{a}} \frac{dk}{2\pi} \sum_{\substack{n_1: \epsilon_{n_1,k}>0,\\
    n_2:\epsilon_{n_2,k}<0}}D(i\omega_n)\left|\braket{u_{n_1,k}|\partial_k u_{n_2,k}}\right|^2\\
    \times[\mathcal{V}_{n_1,n_1}(k)-\mathcal{V}_{n_2,n_2}(k)]\frac{\omega_n^2-(\epsilon_{n_1,k}-\epsilon_{n_2,k})^2}{[\omega_n^2+(\epsilon_{n_1,k}-\epsilon_{n_2,k})^2]^2}\:,
\end{gathered}
\end{equation}

\onecolumngrid
\begin{align} 
\begin{gathered}
    \delta V_{\text{non-diag}}=\frac{g_0^2}{\beta}\sum_{i\omega_n}D(i\omega_n)\int ^{\frac{\pi}{a}}_{-\frac{\pi}{a}}\frac{dk}{2\pi}\sum_{\substack{n_1: \epsilon_{n_1,k}>0,\\n_2:\epsilon_{n_2,k}<0}}\frac{2(\epsilon_{n_1,k}-\epsilon_{n_2,k})}{\omega_n^2+(\epsilon_{n_1,k}-\epsilon_{n_2,k})^2}\\
    \times\Bigg\{\sum_{n_3,n_3\neq n_1}\frac{1}{\epsilon_{n_1,k}-\epsilon_{n_3,k}}\text{Re}\left[\mathcal{V}_{n_1,n_3}(k)\braket{u_{n_3,k}|\partial_k u_{n_2,k}}\braket{\partial_k u_{n_2,k}|u_{n_1,k}}\right]\\ 
    +\sum_{n_3,n_3\neq n_2}\frac{1}{\epsilon_{n_2,k}-\epsilon_{n_3,k}}\text{Re}\left[\mathcal{V}_{n_3,n_2}(k)\braket{u_{n_1,k}|\partial_k u_{n_3,k}}\braket{\partial_k u_{n_2,k}|u_{n_1,k}}\right]\Bigg\}\\ 
    \end{gathered}\\ 
    \begin{gathered}
    +\frac{g_0^2}{\beta}\sum_{i\omega_n}D(i\omega_n)\int^{\frac{\pi}{a}}_{-\frac{\pi}{a}} \frac{dk}{2\pi} \sum_{\substack{n_1: \epsilon_{n_1,k}>0,\\n_2:\epsilon_{n_2,k}<0,\\n_1\neq n_2}}\frac{1}{\omega_n^2+(\epsilon_{n_1,k}-\epsilon_{n_2,k})^2}\frac{2}{\epsilon_{n_1,k}-\epsilon_{n_2,k}}\\   \times\text{Re}\left[\mathcal{V}_{n_1,n_2}(k)\braket{u_{n_2,k}|\partial_k u_{n_1,k}}\right]\left(\frac{\partial\epsilon_{n_2,k}}{\partial k}-\frac{\partial\epsilon_{n_1,k}}{\partial k}\right)\\
    +\frac{g_0^2}{\beta}\sum_{i\omega_n}D(i\omega_n)\int^{\frac{\pi}{a}}_{-\frac{\pi}{a}} \frac{dk}{2\pi} \sum_{\substack{n_1: \epsilon_{n_1,k}>0,\\n_2:\epsilon_{n_2,k}<0,\\n_1\neq n_2}}\frac{2}{\omega_n^2+(\epsilon_{n_1,k}-\epsilon_{n_2,k})^2}\text{Re}\left[\frac{\partial \mathcal{V}_{n_1,n_2}(k)}{\partial k}\braket{u_{n_2,k}|\partial_k u_{n_1,k}}\right]\:, \nonumber
    \end{gathered}
\end{align}
where $\mathcal{V}_{n,n'}(k)$ are matrix element of the operator $\mathbfcal{V}(k)$ in the Bloch wave basis.\\
\twocolumngrid

For a system with two symmetric bands ($\epsilon_{+,k}=-\epsilon_{-,k}\equiv\epsilon_{k}$) the equation above reduces to

\begin{equation}
\begin{gathered}
    \delta V=g_0^2\int^{\frac{\pi}{a}}_{-\frac{\pi}{a}} \frac{dk}{2\pi}\: (\mathcal{V}_{+,+}(k)-\mathcal{V}_{-,-}(k))|\braket{u_{+,k}|\partial_k u_{-,k}}|^2\\
    \times\frac{1}{\beta}\sum_{i\omega_n} D(i\omega_n) \frac{(\omega_n^2-4\epsilon_k^2)}{(\omega_n^2+4\epsilon_k^2)^2}\\
        +2g_0^2\int^{\frac{\pi}{a}}_{-\frac{\pi}{a}} \frac{dk}{2\pi}\:\text{Re}\big[\mathcal{V}_{+,-}(k)(\braket{u_{+,k}|\partial_k u_{+,k}}\\
        -\braket{u_{-,k}|\partial_k u_{-,k}})\braket{ u_{-,k}|\partial_k u_{+,k}}\big]
        \frac{1}{\beta}\sum_{i\omega_n}D(i\omega_n) \frac{1}{\omega_n^2+4\epsilon_k^2}\\
        +2g_0^2\int^{\frac{\pi}{a}}_{-\frac{\pi}{a}} \frac{dk}{2\pi}\:\text{Re}\left[\epsilon_k\frac{\partial}{\partial k}\left(\frac{\mathcal{V}_{+,-}(k)}{\epsilon_k}\right)\braket{u_{-,k}|\partial_k u_{+,k}}\right]\\
        \times\frac{1}{\beta}\sum_{i\omega_n} D(i\omega_n)\frac{1}{\omega_n^2+4\epsilon_k^2}\:. \label{arb_oper}
\end{gathered}
\end{equation}

We use Eq.~\eqref{arb_oper} for the calculation of the cavity-induced charge imbalance correction [see Eq.~\eqref{delta_rho}].

\section{Charge imbalance correction}

The single-particle operator of the charge imbalance operator given by Eq.~\eqref{ChargeImbalance} reads  
\begin{equation}
    \mathbf{\varrho}(k)=\frac{e}{N}\left(\ket{1}\bra{1}-\ket{2}\bra{2}\right)\:,
\end{equation}
where we relabeled basis vectors for convenience.

To calculate the cavity-induced charge imbalance correction [see Eq.~\eqref{delta_rho}], using Eq.~\eqref{arb_oper}, we need to rewrite the one-particle charge imbalance operator $\mathbf{\varrho}$(k) in the Bloch wave basis. For a system with two symmetric bands the basis vectors $\ket{1}$, $\ket{2}$ and the one-particle charge imbalance operator $\mathbf{\varrho}$(k) are written as

\begin{equation}
    \begin{gathered}
        \ket{1}=\braket{u_{+,k}|1}\ket{u_{+,k}}+\braket{u_{-,k}|1}\ket{u_{-,k}}\:,\\
        \ket{2}=\braket{u_{+,k}|2}\ket{u_{+,k}}+\braket{u_{-,k}|2}\ket{u_{-,k}}\:,
    \end{gathered}
\end{equation}

\begin{equation}
\begin{gathered}
    \mathbf{\varrho}(k)=\left(|\braket{u_{+,k}|1}|^2-|\braket{u_{+,k}|2}|^2\right)\ket{u_{+,k}}\bra{u_{+,k}}\\
    +\left(|\braket{u_{-,k}|1}|^2-|\braket{u_{-,k}|2}|^2\right)\ket{u_{-,k}}\bra{u_{-,k}}\\
    +\left(\braket{u_{-,k}|1}\braket{1|u_{+,k}}-\braket{u_-|2}\braket{2|u_{+,k}}\right)\ket{u_{-,k}}\bra{u_{+,k}}\\
    +\left(\braket{u_{+,k}|1}\braket{1|u_{-,k}}-\braket{u_{+,k}|2}\braket{2|u_{-,k}}\right)\ket{u_{+,k}}\bra{u_{-,k}}\;,
\end{gathered}
\end{equation}
and thus
\begin{equation}
\begin{gathered}
\varrho_{+,+}(k)=|\braket{u_{+,k}|1}|^2-|\braket{u_{+,k}|2}|^2\:,\\
\varrho_{-,-}(k)=|\braket{u_{-,k}|1}|^2-|\braket{u_{-,k}|2}|^2\:,\\
\varrho_{+,-}(k)=\braket{u_{+,k}|1}\braket{1|u_{-,k}}-\braket{u_{+,k}|2}\braket{2|u_{-,k}}\:,\\
\varrho_{-,+}(k)=\braket{u_{-,k}|1}\braket{1|u_{+,k}}-\braket{u_{-,k}|2}\braket{2|u_{+,k}}\:. \label{CI_ME}
\end{gathered}
\end{equation}\\

To obtain the cavity-induced charge imbalance correction [see Eq.~$\eqref{delta_rho}$], we substitute expressions for the matrix elements of charge imbalance given by Eq.~\eqref{CI_ME} in Eq.~\eqref{arb_oper}.

\section{Polarization correction (general formula)}
The dressed Green function is approximated as
\begin{equation}
    \mathbf{G}\approx \mathbf{G}_0+\mathbf{G}_1\:,
\end{equation}
where $\mathbf{G}_1=\mathbf{G}_0\mathbf{\Sigma}\mathbf{G}_0$. The polarization correction can then be transformed in the following way:
\onecolumngrid

\begin{align} \label{TransfKotliar}
\begin{gathered}
\frac{\partial P_{\text{cav}}}{\partial\xi} =i \frac{e}{2N}\frac{1}{\beta}\sum_{i\epsilon_m}{\rm Tr}\Bigg\{\frac{\partial \mathbf{G}_{0}^{-1}(k)}{\partial k}\frac{\partial \mathbf{G}_{1}(k)}{\partial i\epsilon_m}\frac{\partial \mathbf{G}_{0}^{-1}(k)}{\partial\xi}\mathbf{G}_{0}(k)
+\frac{\partial \mathbf{G}_{0}^{-1}(k)}{\partial k}\frac{\partial \mathbf{G}_{0}(k)}{\partial i\epsilon_m}\frac{\partial \mathbf{G}_{0}^{-1}(k)}{\partial\xi}\mathbf{G}_{1}(k) \nonumber
  \end{gathered}\\
 \begin{gathered}
-\frac{\partial \mathbf{\Sigma}(k)}{\partial k}\frac{\partial \mathbf{G}_{0}(k)}{\partial i\epsilon_m}\frac{\partial \mathbf{G}_{0}^{-1}(k)}{\partial\xi}\mathbf{G}_{0}(k)
-\frac{\partial \mathbf{G}_{0}^{-1}(k)}{\partial k}\frac{\partial \mathbf{G}_{0}(k)}{\partial i\epsilon_m}\frac{\partial \mathbf{\Sigma}(k)}{\partial\xi}\mathbf{G}_{0}(k)
 -\frac{\partial \mathbf{G}_{0}^{-1}(k)}{\partial\xi}\frac{\partial \mathbf{G}_{1}(k)}{\partial i\epsilon_m}\frac{\partial \mathbf{G}_{0}^{-1}(k)}{\partial k}\mathbf{G}_{0}(k)\\ 
 -\frac{\partial \mathbf{G}_{0}^{-1}(k)}{\partial\xi}\frac{\partial \mathbf{G}_{0}(k)}{\partial i\epsilon_m}\frac{\partial \mathbf{G}_{0}^{-1}(k)}{\partial k}\mathbf{G}_{1}(k)
 +\frac{\partial \mathbf{\Sigma}(k)}{\partial\xi}\frac{\partial \mathbf{G}_{0}(k)}{\partial i\epsilon_m}\frac{\partial \mathbf{G}_{0}^{-1}(k)}{\partial k}\mathbf{G}_{0}(k)
 +\frac{\partial \mathbf{G}_{0}^{-1}(k)}{\partial\xi}\frac{\partial \mathbf{G}_{0}(k)}{\partial i\epsilon_m}\frac{\partial \mathbf{\Sigma}(k)}{\partial k}\mathbf{G}_{0}(k)\Bigg\}\\
 =i \frac{e}{2N}\frac{1}{\beta}\sum_{i\epsilon_m}{\rm Tr}\Bigg\{\frac{\partial \mathbf{G}_{0}(k)}{\partial k} \mathbf{\Sigma}(k)\frac{\partial \mathbf{G}_{0}(k)}{\partial\xi}
 -\frac{\partial \mathbf{G}_{0}(k)}{\partial k}\frac{\partial \mathbf{G}_{0}(k)}{\partial\xi} \mathbf{\Sigma}(k)-\frac{\partial \mathbf{\Sigma}(k)}{\partial k}\mathbf{G}_{0}(k)\frac{\partial \mathbf{G}_{0}(k)}{\partial\xi}
 -\frac{\partial \mathbf{G}_{0}(k)}{\partial k}\mathbf{G}_{0}(k)\frac{\partial \mathbf{\Sigma}(k)}{\partial\xi}\\
-\frac{\partial \mathbf{G}_{0}(k)}{\partial\xi}\mathbf{\Sigma}(k)\frac{\partial \mathbf{G}_{0}(k)}{\partial k}+\frac{\partial \mathbf{G}_{0}(k)}{\partial\xi}\frac{\partial \mathbf{G}_{0}(k)}{\partial k}\mathbf{\Sigma}(k)
+\frac{\partial \mathbf{\Sigma}(k)}{\partial\xi}\mathbf{G}_{0}(k)\frac{\partial \mathbf{G}_{0}(k)}{\partial k}+\frac{\partial \mathbf{G}_{0}(k)}{\partial\xi}\mathbf{G}_{0}(k)\frac{\partial \mathbf{\Sigma}(k)}{\partial k}\Bigg\}\\
 =i \frac{e}{2N}\frac{1}{\beta}\sum_{i\epsilon_m}{\rm Tr}\Bigg\{-\frac{\partial^{2}\mathbf{G}_{0}(k)}{\partial\xi\partial k}\mathbf{G}_{0}(k)\mathbf{\Sigma}(k)-\frac{\partial \mathbf{G}_{0}(k)}{\partial k}\frac{\partial \mathbf{G}_{0}(k)}{\partial\xi}\mathbf{\Sigma}(k)
 -\frac{\partial \mathbf{G}_{0}(k)}{\partial k}\mathbf{G}_{0}(k)\frac{\partial \mathbf{\Sigma}(k)}{\partial\xi} +\mathbf{G}_{0}(k)\frac{\partial^{2}\mathbf{G}_{0}(k)}{\partial\xi\partial k}\mathbf{\Sigma}(k)\\
 +\frac{\partial \mathbf{G}_{0}(k)}{\partial\xi}\frac{\partial \mathbf{G}_{0}(k)}{\partial k}\mathbf{\Sigma}(k)+\mathbf{G}_{0}(k)\frac{\partial \mathbf{G}_{0}(k)}{\partial k}\frac{\partial \mathbf{\Sigma}(k)}{\partial\xi}\Bigg\}
 =i \frac{e}{2N}\frac{\partial}{\partial\xi}\frac{1}{\beta}\sum_{i\epsilon_m}{\rm Tr}\left\{ \left(\mathbf{G}_{0}(k)\frac{\partial \mathbf{G}_{0}(k)}{\partial k}-\frac{\partial \mathbf{G}_{0}(k)}{\partial k}\mathbf{G}_{0}(k)\right)\mathbf{\Sigma}(k)\right\}\:, 
\end{gathered}
\end{align}
where $\mathbf{\Sigma} (k)$ is the diagonal in $k$ part of the self-energy $\mathbf{\Sigma}$.\\
\twocolumngrid

In the Eq.~\eqref{TransfKotliar} on the first step we (i) moved in the 1st and in the 5th terms 
the derivative with respect to $i\epsilon$ from $\mathbf{G}_{1}$ to $\mathbf{G}_{0}$, and (ii) used the
identities $\partial_{i\epsilon_m}\mathbf{G}_{0}=-\mathbf{G}_{0}\mathbf{G}_{0}$ and $\mathbf{G}_{0}(\partial \mathbf{G}_{0}^{-1})\mathbf{G}_{0}=-\partial \mathbf{G}_{0}$, where $\partial$ is a partial derivative with respect to any variable.
In the second step, we performed a partial $k$-integration in the
1st and in the 5th term, and then used the cyclic invariance of the trace
to move $\mathbf{\Sigma}$ to the same position in all terms. As a result, Eq.~\eqref{TransfKotliar}
reduces to a total derivative with respect to the adiabatic parameter
\begin{equation}
\begin{gathered}
\frac{\partial P_{\text{cav}}}{\partial\xi}=i\frac{e}{2N}\frac{\partial}{\partial\xi}\frac{1}{\beta}\sum_{i\epsilon_m}{\rm Tr}\Bigg\{ \Bigg(\mathbf{G}_{0}(k)\frac{\partial \mathbf{G}_{0}(k)}{\partial k}\\
-\frac{\partial \mathbf{G}_{0}(k)}{\partial k}\mathbf{G}_{0}(k)\Bigg)\mathbf{\Sigma}(k)\Bigg\} \\
=i\frac{e}{2N}\frac{\partial}{\partial\xi}\frac{1}{\beta}\sum_{i\epsilon_m}{\rm Tr}\left\{ \left[\mathbf{G}_{0}(k),\frac{\partial \mathbf{G}_{0}(k)}{\partial k}\right]\mathbf{\Sigma}(k)\right\}, \label{dP/dx-1}
\end{gathered}
\end{equation}
as it was announced in Eq.~\eqref{dP/dx-final}.
\section{Polarization correction for a dielectric with two symmetric bands embedded in a cavity}
\begin{figure}[b]
\includegraphics[width=0.2\textwidth]{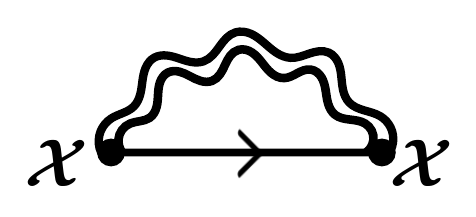}
\caption{Diagrammatic representation of the lowest-order self-energy $\mathbf{\Sigma}$. The analytical expression for the diagram is given in Eq.~\eqref{eq:Self-energy}. Here, $\mathbfcal{X}$ is the one-particle position operator given by Eq.~\eqref{CO_OneParticle}. Solid lines denote bare electron Green functions, see Eq.~\eqref{BareGF}. The double wavy line is a dressed photon propagator in the RPA given by Eq.~\eqref{DressedPP}. } \label{Self-energy}
\end{figure}

Cavity-induced polarization correction is written as
\begin{equation}
\begin{gathered}
    \delta P_{\text{cav}}= i\frac{e}{2N}\frac{1}{\beta}\sum_{i\epsilon_m}{\rm Tr}\left\{ \left[\mathbf{G}_{0}(k),\frac{\partial \mathbf{G}_{0}(k)}{\partial k}\right]\mathbf{\Sigma}(k)\right\}(\xi=\Delta)\\
    -i\frac{e}{2N}\frac{1}{\beta}\sum_{i\epsilon_m}{\rm Tr}\left\{ \left[\mathbf{G}_{0}(k),\frac{\partial \mathbf{G}_{0}(k)}{\partial k}\right]\mathbf{\Sigma}(k)\right\}(\xi=0)\:, \label{PCfinal}
\end{gathered}
\end{equation}
where $\mathbf{\Sigma} (k)$ is the diagonal in $k$ part of the lowest-order self-energy $\mathbf{\Sigma}$ given by 
\begin{equation} \label{eq:Self-energy}
    \mathbf{\Sigma}(i\epsilon_m)=(-1)\cdot\frac{g_0^2}{N\beta}\sum_{i\omega_n} D(i\omega_n)\mathbfcal{X}\mathbf{G}_0(i\epsilon_m+i\omega_n)\mathbfcal{X}\:.
\end{equation}
The diagrammatic representation of the equation above is shown in Fig.~\ref{Self-energy}.

Taking into account that the bare electron Green function can be represented as
\begin{equation}   \textbf{G}_0(k)=\sum_n\frac{\ket{u_{n,k}}\bra{u_{n,k}}}{i\epsilon_m-\epsilon_{n,k}}\:, \label{BareGF}
\end{equation}
parts of the Eq.~\eqref{PCfinal} can be rewritten in the following way (we use the fact that $\partial_{k}\langle u_{n,k}|u_{n',k}\rangle=0$):
\begin{align}
\begin{gathered}
    \mathbf{G}_0(k)\frac{\partial \mathbf{G}_0(k)}{\partial k}-\frac{\partial \mathbf{G}_0(k)}{\partial k}\mathbf{G}_0(k)\\
    =\sum_{n_1,n_2}\ket{u_{n_1k}}\braket{u_{n_1,k}|\partial_k u_{n_2,k}}\bra{u_{n_2,k}}\times \\ \nonumber
\end{gathered}\\ 
\begin{gathered}
    \times\Bigg(\frac{2}{(i\epsilon_m-\epsilon_{n_1,k})(i\epsilon_m-\epsilon_{n_2,k})}\\
    -\frac{1}{(i\epsilon_m-\epsilon_{n_1,k})^2}
    -\frac{1}{(i\epsilon_m-\epsilon_{n_2,k})^2}\Bigg)\:,
\end{gathered}
\end{align}
\begin{equation}
    \begin{gathered}
    \mathbf{\Sigma}=\frac{g_0^2}{N}\sum_{n_1,n_3}\sum_{k_1,k_3}\ket{u_{n_1,k_1}}\bra{u_{n_3,k_3}} 
    \sum_{n_2}\sum_{k_2}\frac{1}{\beta}\sum_{i\omega_n}\frac{D(i\omega_n)}{2q^2}\\
   \times\left[\delta_{k_2,k_1+q}-\delta_{k_2,k_1-q}\right]
    \braket{u_{n_1,k_1}|u_{n_2,k_2}}
    \frac{1}{i(\epsilon_m+\omega_n)-\epsilon_{n_2,k_2}}\\
    \times\left[\delta_{k_3,k_2+q}-\delta_{k_3,k_2-q}\right]
     \braket{u_{n_2,k_2}|u_{n_3,k_3}}\\
     =\sum_{n_1,n_3}\sum_{k_1,k_2}\ket{u_{n_1,k_1}}\bra{u_{n_3,k_2}}
   \sum_{n_2}\frac{1}{\beta}\sum_{i\omega_n}\frac{D(i\omega_n)}{2q^2}\\
    \times\Bigg\{\frac{1}{i(\epsilon_m+\omega_n)-\epsilon_{n_2,k_1+q}}\braket{u_{n_1,k_1}|u_{n_2,k_1+q}}\\
    \times\braket{u_{n_2,k_1+q}|u_{n_3,k_2}}
    \left(\delta_{k_2,k_1+2q}-\delta_{k_2,k_1}\right)
    +(q\to-q)\Bigg\}\:.
    \end{gathered}
\end{equation}

We choose only diagonal in $k$-terms [as we need only them for Eq.~\eqref{PCfinal}],
taking into account that $q$ is finite, and, as a result, we obtain 
\begin{equation}
\begin{gathered}
\mathbf{\Sigma}(k)=-\frac{g_0^2}{N}\sum_{n_1,n_3}\sum_{k}\ket{u_{n_1,k}}\bra{u_{n_3,k}}\sum_{n_2}\frac{1}{\beta}\sum_{i\omega_n}\frac{D(i\omega_n)}{2q^2}\\
\times\Bigg\{\frac{1}{i(\epsilon_m+\omega_n)-\epsilon_{n_2,k+q}}\braket{u_{n_1,k}|u_{n_2,k+q}}\\
\times\braket{u_{n_2,k+q}|u_{n_3,k}}
    +(q\to-q)\Bigg\}.
\end{gathered}
\end{equation}
After the summation over the Matsubara frequency $i\epsilon_m$, taking the limit $q\to 0$ for the dielectric with two symmetric bands embedded in a cavity in the case of the zero temperature we obtain the following expression
\begin{equation}
\begin{gathered}
    \delta P_{\text{cav}}= -i\frac{e}{2N}g_0^2\frac{1}{\beta}\sum_{i\omega_n}\int^{\frac{\pi}{a}}_{-\frac{\pi}{a}} \frac{dk}{2\pi}\: D(i\omega_n)\frac{\omega_n^2+12\epsilon_k^2}{(\omega_n^2+4\epsilon_k^2)^2}\\
    \times\bigg\{\braket{u_{+,k}|\partial_k u_{-,k}}\braket{u_{-,k}|\partial^2_k u_{+,k}}\\
    -\braket{u_{-,k}|\partial_k u_{+,k}}\braket{u_{+,k}|\partial^2_k u_{-,k}}+2|\braket{u_{-,k}|\partial_k u_{+,k}}|^2\\    \times\left(\braket{u_{+,k}|\partial_ku_{+,k}}-\braket{u_{-,k}|\partial_ku_{-,k}}\right)\bigg\}\:,
\end{gathered}
\end{equation}
which we use for the calculation of the cavity-induced polarization correction for the Rice-Mele model embedded in a cavity in Sec.~IV~B.

\section{Relation between charge imbalance and current}
The time derivative of the charge imbalance operator given by Eq.~\eqref{ChargeImbalance} is written as
\begin{equation}
\begin{gathered}
    \frac{d\hat{\rho}}{dt}=\frac{e}{N}\sum_k \Bigg\{\frac{d\hat{b}^{\dagger}_k}{dt} \hat{b}_k+\hat{b}^{\dagger}_k \frac{d\hat{b}_k}{dt}\\
    -\left(\frac{d\hat{c}^{\dagger}_k}{dt}\hat{c}_k+\hat{c}_k^{\dagger}\frac{d\hat{c}_k}{dt}\right)\Bigg\}\:.
\end{gathered}
\end{equation}
The evolution of annihilation operators $\hat{b}_k$ and $\hat{c}_k$ is found from the Heisenberg equation of motion,
\begin{equation}
    \begin{gathered}
    \frac{d\hat{b}_k}{dt}=i\left[
    \hat{H},\hat{b}_k\right]\:,\\
    \frac{d\hat{c}_k}{dt}=i\left[ \hat{H},\hat{c}_k\right]\:.
    \end{gathered}
\end{equation}
For the Rice-Mele model with the Hamiltonian given by Eq.~\eqref{Rice-MeleHamiltonian}, 
the evolution of annihilation operators $\hat{b}_k$ and $\hat{c}_k$ is described by the following equations
\begin{equation}
\begin{gathered}
    \frac{d\hat{b}_k}{dt}=i\left\{-\Delta \hat{b}_k - \hat{c}_k\left(t_1 e^{ikd_1}+t_2e^{-ikd_2}\right)\right\}\:,\\
    \frac{d\hat{c}_k}{dt}=i\left\{\Delta \hat{c}_k-\hat{b}_k\left(t_1 e^{-ikd_1}+t_2e^{ikd_2}\right)\right\}\:,
\end{gathered}
\end{equation}
and, therefore, the time derivative of the charge imbalance is given by

\begin{equation}
\begin{gathered}
    \frac{d\hat{\rho}}{dt}=\frac{2ie}{N}\sum_k\Big\{\hat{c}^{\dagger}_k\hat{b}_k\left(t_1e^{-ikd_1}+t_2e^{ikd_2}\right)\\
    -\hat{b}_k^{\dagger}\hat{c}_k\left(t_1 e^{ikd_1}+t_2e^{-ikd_2}\right)\Big\}\:.
\end{gathered}   
\end{equation}
And in the limit of the separate dimers ($t_2=0$), the equation above reduces to

\begin{equation}
    \frac{d\hat{\rho}}{dt}=\frac{2ie}{N}\sum_k\left\{\hat{c}^{\dagger}_k\hat{b}_kt_1e^{-ikd_1}-\hat{b}_k^{\dagger}\hat{c}_k t_1e^{ikd_1}\right\}\:. \label{ChImb}    
\end{equation}
On the other hand, the current operator, by definition, is
\begin{equation}
    \hat{J}=\frac{e}{L}\sum_k\hat{\psi}_k^{\dagger} \frac{\partial \mathbf{h}(k)}{\partial k} \hat{\psi}_k\:.
\end{equation}
Thus, for the Rice-Mele model in the case of separate dimers, the current is given by 
\begin{equation}
    \hat{J}=-ied_1\sum_k\left\{\hat{c}^{\dagger}_k\hat{b}_kt_1e^{-ikd_1}-\hat{b}_k^{\dagger}\hat{c}_k t_1e^{ikd_1}\right\}\:.\label{Curr}
\end{equation}
Comparing Eqs.~\eqref{ChImb} -- \eqref{Curr} in the case of  equally spaced sites ($d_1=d_2=a/2$), we obtain the 
continuity equation, 
\begin{equation}
    \frac{d\hat{\rho}}{dt}=-4\hat{J}\:,
\end{equation}
as expected.
\twocolumngrid

\bibliography{bibliography}

\begin{thebibliography}{67}
\bibitem [{not()}]{note1overN}%
  \BibitemOpen
  \href@noop {} {}\bibinfo {note} {The "exactness of RPA" at large $N$
  implicitly assumes that in the formal limit $N\to\infty$ the number of
  relevant photon modes remains finite. Physically, this assumes that the field
  is always strongly confined inside the cavity. Also, if the direct (Coulomb)
  interaction between the electrons becomes relevant, the notion of "cavity
  RPA" that becomes exact in the large $N$ limit has to be refined. Apparently,
  the $1/N$ argument rules out only the vertex corrections due to the
  cavity-photon-mediated interaction. Hence, the cavity RPA diagrams dominating
  in the large N limit correspond to the RPA-like bubbles as in Fig. 2 where
  the bubbles stand for the polarizability of the interacting electronic
  system. In this work, we are not interested in the effect of the direct
  interaction.}\BibitemShut {Stop}%
\end{thebibliography}

\end{document}